\documentclass[aps,prd,twocolumn,groupedaddress,floatfix,superscriptaddress,showpacs,showkeys]{revtex4-1}
\usepackage{graphicx}   % for including figures
\usepackage[caption=false]{subfig}
\usepackage{amsmath}

\begin{document}
%\linenumbers
\title{A Search for UHE Tau Neutrinos with IceCube}
%\email[]{seo@physto.se}
%\thanks{}
%\affiliation{Stockholm University}

%\collaboration{IceCube}

\affiliation{III. Physikalisches Institut, RWTH Aachen University, D-52056 Aachen, Germany}
\affiliation{School of Chemistry \& Physics, University of Adelaide, Adelaide SA, 5005 Australia}
\affiliation{Dept.~of Physics and Astronomy, University of Alaska Anchorage, 3211 Providence Dr., Anchorage, AK 99508, USA}
\affiliation{CTSPS, Clark-Atlanta University, Atlanta, GA 30314, USA}
\affiliation{School of Physics and Center for Relativistic Astrophysics, Georgia Institute of Technology, Atlanta, GA 30332, USA}
\affiliation{Dept.~of Physics, Southern University, Baton Rouge, LA 70813, USA}
\affiliation{Dept.~of Physics, University of California, Berkeley, CA 94720, USA}
\affiliation{Lawrence Berkeley National Laboratory, Berkeley, CA 94720, USA}
\affiliation{Institut f\"ur Physik, Humboldt-Universit\"at zu Berlin, D-12489 Berlin, Germany}
\affiliation{Fakult\"at f\"ur Physik \& Astronomie, Ruhr-Universit\"at Bochum, D-44780 Bochum, Germany}
\affiliation{Physikalisches Institut, Universit\"at Bonn, Nussallee 12, D-53115 Bonn, Germany}
\affiliation{Dept.~of Physics, University of the West Indies, Cave Hill Campus, Bridgetown BB11000, Barbados}
\affiliation{Universit\'e Libre de Bruxelles, Science Faculty CP230, B-1050 Brussels, Belgium}
\affiliation{Vrije Universiteit Brussel, Dienst ELEM, B-1050 Brussels, Belgium}
\affiliation{Dept.~of Physics, Chiba University, Chiba 263-8522, Japan}
\affiliation{Dept.~of Physics and Astronomy, University of Canterbury, Private Bag 4800, Christchurch, New Zealand}
\affiliation{Dept.~of Physics, University of Maryland, College Park, MD 20742, USA}
\affiliation{Dept.~of Physics and Center for Cosmology and Astro-Particle Physics, Ohio State University, Columbus, OH 43210, USA}
\affiliation{Dept.~of Astronomy, Ohio State University, Columbus, OH 43210, USA}
\affiliation{Dept.~of Physics, TU Dortmund University, D-44221 Dortmund, Germany}
\affiliation{Dept.~of Physics, University of Alberta, Edmonton, Alberta, Canada T6G 2G7}
\affiliation{D\'epartement de physique nucl\'eaire et corpusculaire, Universit\'e de Gen\`eve, CH-1211 Gen\`eve, Switzerland}
\affiliation{Dept.~of Physics and Astronomy, University of Gent, B-9000 Gent, Belgium}
\affiliation{Dept.~of Physics and Astronomy, University of California, Irvine, CA 92697, USA}
\affiliation{Laboratory for High Energy Physics, \'Ecole Polytechnique F\'ed\'erale, CH-1015 Lausanne, Switzerland}
\affiliation{Dept.~of Physics and Astronomy, University of Kansas, Lawrence, KS 66045, USA}
\affiliation{Dept.~of Astronomy, University of Wisconsin, Madison, WI 53706, USA}
\affiliation{Dept.~of Physics, University of Wisconsin, Madison, WI 53706, USA}
\affiliation{Institute of Physics, University of Mainz, Staudinger Weg 7, D-55099 Mainz, Germany}
\affiliation{Universit\'e de Mons, 7000 Mons, Belgium}
\affiliation{T.U. Munich, D-85748 Garching, Germany}
\affiliation{Bartol Research Institute and Department of Physics and Astronomy, University of Delaware, Newark, DE 19716, USA}
\affiliation{Dept.~of Physics, University of Oxford, 1 Keble Road, Oxford OX1 3NP, UK}
\affiliation{Dept.~of Physics, University of Wisconsin, River Falls, WI 54022, USA}
\affiliation{Oskar Klein Centre and Dept.~of Physics, Stockholm University, SE-10691 Stockholm, Sweden}
\affiliation{Department of Physics and Astronomy, Stony Brook University, Stony Brook, NY 11794-3800, USA}
\affiliation{Dept.~of Physics and Astronomy, University of Alabama, Tuscaloosa, AL 35487, USA}
\affiliation{Dept.~of Astronomy and Astrophysics, Pennsylvania State University, University Park, PA 16802, USA}
\affiliation{Dept.~of Physics, Pennsylvania State University, University Park, PA 16802, USA}
\affiliation{Dept.~of Physics and Astronomy, Uppsala University, Box 516, S-75120 Uppsala, Sweden}
\affiliation{Dept.~of Physics, University of Wuppertal, D-42119 Wuppertal, Germany}
\affiliation{DESY, D-15735 Zeuthen, Germany}

\author{R.~Abbasi}
\affiliation{Dept.~of Physics, University of Wisconsin, Madison, WI 53706, USA}
\author{Y.~Abdou}
\affiliation{Dept.~of Physics and Astronomy, University of Gent, B-9000 Gent, Belgium}
\author{T.~Abu-Zayyad}
\affiliation{Dept.~of Physics, University of Wisconsin, River Falls, WI 54022, USA}
\author{M.~Ackermann}
\affiliation{DESY, D-15735 Zeuthen, Germany}
\author{J.~Adams}
\affiliation{Dept.~of Physics and Astronomy, University of Canterbury, Private Bag 4800, Christchurch, New Zealand}
\author{J.~A.~Aguilar}
\affiliation{D\'epartement de physique nucl\'eaire et corpusculaire, Universit\'e de Gen\`eve, CH-1211 Gen\`eve, Switzerland}
\author{M.~Ahlers}
\affiliation{Dept.~of Physics, University of Wisconsin, Madison, WI 53706, USA}
\author{D.~Altmann}
\affiliation{III. Physikalisches Institut, RWTH Aachen University, D-52056 Aachen, Germany}
\author{K.~Andeen}
\affiliation{Dept.~of Physics, University of Wisconsin, Madison, WI 53706, USA}
\author{J.~Auffenberg}
\affiliation{Dept.~of Physics, University of Wisconsin, Madison, WI 53706, USA}
% ...

\date{\today}
\collaboration{IceCube Collaboration}
\noaffiliation

%
% A few definitions to make life easier
% A trailing "NS" means "No Space" is added after.
%
\newcommand{\Enu}{$\rm{E}_\nu$}
\newcommand{\nue}{$\nu_{\rm e}\;$}
\newcommand{\numu}{$\nu_\mu\;$}
\newcommand{\numubar}{$\overline{\nu_\mu}\;$}
\newcommand{\nutau}{$\nu_\tau\;$}
\newcommand{\lqden}{$\rho_q\;$}
\newcommand{\lqdenI}{$\rho_q(I)\;$}
\newcommand{\lqdenII}{$\rho_q(II)\;$}
\newcommand{\lqdenIII}{$\rho_q(III)\;$}
\newcommand{\irmax}{IR$_{\rm max}\;$}
\newcommand{\lnpe}{$\log_{10}{\rm N_{pe}}\;$}
\newcommand{\npe}{${\rm N_{\rm pe}}\;$}
\newcommand{\Aeff}{${\rm A}_{\rm eff}\;$}
\newcommand{\NpeDOM}{${\rm N_{\rm pe}^{\rm DOM}}\;$}
\newcommand{\NpeDOMvsT}{${\rm N_{\rm pe}^{\rm DOM}(t)}\;$}
\newcommand{\nueNS}{$\nu_{\rm e}$}
\newcommand{\numuNS}{$\nu_\mu$}
\newcommand{\numubarNS}{$\overline{\nu_\mu}$}
\newcommand{\nutauNS}{$\nu_\tau$}
\newcommand{\lqdenNS}{$\rho_q$}
\newcommand{\lqdenINS}{$\rho_q(I)$}
\newcommand{\lqdenIINS}{$\rho_q(II)$}
\newcommand{\lqdenIIINS}{$\rho_q(III)$}
\newcommand{\irmaxNS}{IR$_{\rm max}$}
\newcommand{\lnpeNS}{$\log_{10}{\rm N_{pe}}$}
\newcommand{\npeNS}{${\rm N_{\rm pe}}$}
\newcommand{\AeffNS}{${\rm A}_{\rm eff}$}
\newcommand{\NpeDOMvsTNS}{${\rm N_{\rm pe}^{\rm DOM}(t)}$}

%%--- March 13, 2012 update (subtrace 3 background events in the limit)
%--- All Flavor limit: 0.60  BG events, 3 observed events, (340 TeV - 200 PeV)
%\newcommand{\theLimit}{$E^{2} \Phi(\nu_{\rm x}) < 2.92 \times 10^{-8}\, {\rm GeV}\,{\rm cm}^{-2}\, {\rm sr}^{-1}\, {\rm s}^{-1}\;$}

%%---below: Dec. 8, 2011
%--- All Flavor limit: 0.60  BG events, 3 observed events, (340 TeV - 200 PeV)
%    use this conservative limit for the publication
\newcommand{\theLimit}{$E_{\nu}^{2} \Phi_{90}(\nu_{\rm x}) < 16.3 \times 10^{-8}\, {\rm GeV}\,{\rm cm}^{-2}\, {\rm sr}^{-1}\, {\rm s}^{-1}\;$}

\newcommand{\theTauLimit}{$E^{2} \Phi(\nu_{\rm tau}) < 5.4 \times 10^{-8}\, {\rm GeV}\,{\rm cm}^{-2}\, {\rm sr}^{-1}\, {\rm s}^{-1}\;$}

%--- All Flavor limit: 0.60  BG events, 2 observed events, (340 TeV - 200 PeV) 
\newcommand{\theLimitTwo}{$E^{2} \Phi(\nu_{\rm x}) < 9.0 \times 10^{-8}\, {\rm GeV}\,{\rm cm}^{-2}\, {\rm sr}^{-1}\, {\rm s}^{-1}\;$}

%--- Tau limit: 0.60  BG events, 2 observed events, (331 TeV - 166 PeV) 
\newcommand{\theTauLimitTwo}{$E^{2} \Phi(\nu_{\rm tau}) < 3.0 \times 10^{-8}\, {\rm GeV}\,{\rm cm}^{-2}\, {\rm sr}^{-1}\, {\rm s}^{-1}\;$}

\newcommand{\theAllSens}{$E^{2} \Phi(\nu_{\rm tau}) < 8.6 \times 10^{-8}\, {\rm GeV}\,{\rm cm}^{-2}\, {\rm sr}^{-1}\, {\rm s}^{-1}\;$}
\newcommand{\theTauSens}{$E^{2} \Phi(\nu_{\rm tau}) < 2.87 \times 10^{-8}\, {\rm GeV}\,{\rm cm}^{-2}\, {\rm sr}^{-1}\, {\rm s}^{-1}\;$}

\begin{abstract}
  The first dedicated search for ultra-high energy (UHE) tau neutrinos
  of astrophysical origin was performed using the IceCube detector in
  its 22-string configuration with an instrumented volume of roughly
  0.25~km$^3$.  The search also had sensitivity to UHE electron and
  muon neutrinos.  After application of all selection criteria to
  approximately 200 live-days of data, we expect a background of $0.60
  \pm 0.19$ (stat.) $^{+0.56}_{-0.58}$ (syst.)  events and observe
  three events, which after inspection emerge as being compatible with
  background but are kept in the final sample.  Therefore, we set an
  upper limit on neutrinos of all-flavors from UHE astrophysical
  sources at 90$\%$~CL of \theLimit over an estimated primary neutrino
  energy range of 340~TeV to 200~PeV.
\end{abstract}

% insert suggested PACS numbers in braces on next line \pacs{}

\maketitle

%       Introduction 
%---------------------------------------
\section{Introduction}
\label{sec:Introduction}

Proposed astrophysical sources of observed ultra-high energy (UHE)
cosmic rays are expected to also produce ultra-high energy neutrinos,
mainly via charged pion decay following interactions on ambient matter
and radiation~\cite{HalzenHooper,Becker}. Candidate neutrino sources
include active galactic nuclei, gamma ray bursts and
microquasars~\cite{SteckerDoneSalomonSommers,WB,LevinsonWaxman}.
Neutrinos are expected to arrive at Earth with a flavor ratio of
\nueNS:\numuNS:\nutau = 1:1:1 in the standard neutrino oscillation
scenario~\cite{LearnedPakvasa}.  Other neutrino production and
propagation models predict different flux ratios at
Earth~\cite{RachenMezaros, KashtiWaxman, Kachelriess}.
If there are many astrophysical point sources of neutrinos, but each
one is too weak to be distinguished individually from background, then
a suitable detection strategy is to perform a cumulative search for
``diffuse'' flux of UHE neutrinos over the full available solid angle.

In previous
searches~\cite{DiffuseNumuSearch,DiffuseCascadeSearch1,DiffuseCascadeSearch2},
diffuse astrophysical UHE neutrinos were distinguished from
atmospheric neutrinos by requiring the energy of candidate UHE
neutrino events to exceed a certain threshold.  In this work, we
present techniques for identifying \nutau interactions and show the
results of the first search for diffuse astrophysical UHE neutrinos
that specifically selected events consistent with several \nutau
interaction topologies.

\begin{figure}[h]
\includegraphics[width=70mm]{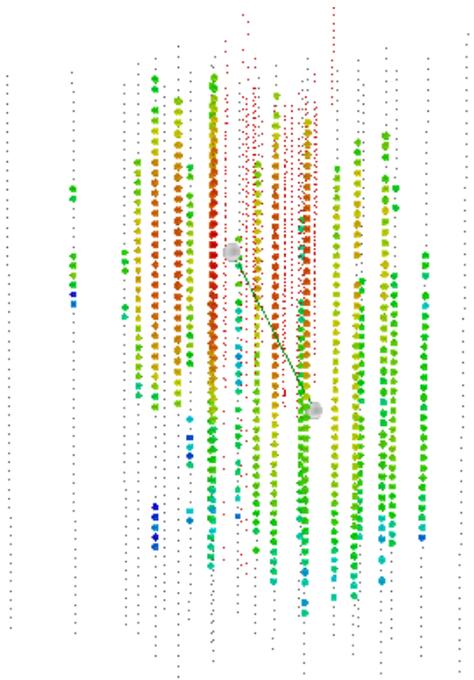}
\caption{\label{fig:db-view} (Color online) A simulated double bang
  event with a primary $\nu_{\tau}$ energy of 47 PeV entering the IC22
  detector at a zenith angle of 35$^{\circ}$.  The two bigger circles
  in gray color represent the vertexes of tau creation (upper left)
  and decay (lower right) which are connected by a tau track.  Each
  smaller circle represents a photomultiplier tube (PMT) that observed
  light produced by the event.  The denser PMTs in the upper middle
  belong to AMANDA and not used in this analysis.  The arrival times
  are indicated by colors that are ordered following the colors of the
  rainbow, with red corresponding to earlier times and violet to later
  times.}
\end{figure}

At $\rm{E}_\nu \gtrsim 1$~PeV, a search for UHE \nutau offers several
advantages over UHE \nue and \numu searches.  Partially and fully
contained interactions of UHE \nutau in the detector can produce very
distinctive signatures owing to the macroscopic $\tau$ decay length.
Each such signature should appear in proportion to the well-measured
$\tau$ branching ratios~\cite{PDG}, providing a useful cross-check on
the positive identification of multiple \nutau events.  As shown
below, the chief sources of possible background events are unlikely to
mimic these signatures.  Also, at these energies there is negligible
intrinsic \nutau background in the conventional atmospheric neutrino
flux~\cite{Bartol}.  The prompt \nutau flux from charm hadron decays
in cosmic-ray-induced air showers is also expected to be
small~\cite{promptNu-1,promptNu-2,promptNu-3}.  The majority of the signal \nutau is
expected to come from the vicinity of the horizon since there is
insufficient material for interactions in the downward-going direction
and \nutau passing through the Earth emerge~\cite{Halzen-1998} at
energies too low to create a UHE signature.

The \nutau event topology depends on how much of the event is
contained in the detector, the \nutau energy, and the composition of
the $\tau$ decay products.  In this work only non-muonic $\tau$ decays
were considered.  A partially contained UHE \nutau having only the
decay vertex of $\tau$ in the instrumented volume is denoted a
``lollipop,'' while one having only the production vertex of the
$\tau$ in the instrumented volume is denoted an ``inverted lollipop.''
A fully contained UHE \nutau having both production and decay vertices
well separated in the instrumented volume is denoted a ``double
bang''~\cite{Learned-1995}.  Fig.~\ref{fig:db-view} shows a simulated
double bang event in the 22-string configuration of the IceCube
detector (IC22) which had an instrumented volume of roughly
0.25~km$^3$.

Applying criteria to identify lollipop, inverted lollipop and double
bang signatures produced by \nutau interactions, we derived limits on
the diffuse UHE neutrino flux.  We assumed a flux ratio of
\nueNS:\numuNS:\nutau = 1:1:1 for this analysis.  We used 282.4
live-days of data collected in 2007-2008 by IC22.  We describe the
IC22 detector in Section~\ref{sec:detector} and the experimental and
simulated data samples in Section~\ref{sec:data}.  We present our
analysis in Section~\ref{sec:nutauID} and the results in
Section~\ref{sec:results}.  We discuss systematic errors in
Section~\ref{sec:systematics} and our conclusions in
Section~\ref{sec:conclusions}.

%----------------------------
%        The Detector
%----------------------------
\section{The IceCube 22-String Detector}
\label{sec:detector}

\begin{figure}[h]
\includegraphics[width=86mm]{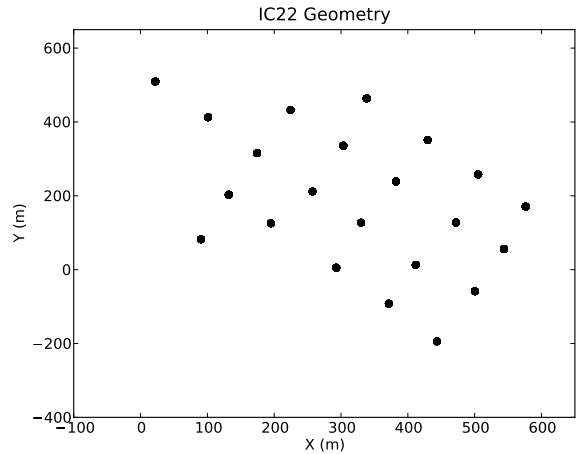}
\caption{\label{fig:ic22-geo} Top view of the IceCube 22 string detector.
Each string is represented as a dot.}
\end{figure}

The 22-string configuration of IceCube (IC22) was deployed in early
2007, began taking physics-quality data in May of that year, and ended
at the transition to IceCube's 40-string configuration in April
2008. Each string consists of 60 digital optical modules (DOMs) buried
deep in the icecap at the South Pole, with regular 17~m vertical
spacing from 1450 to 2450~m below the surface, for a total of 1320
DOMs.  The strings are situated on a regular grid with 125~m
horizontal interstring spacing, covering the area shown in
Fig.~\ref{fig:ic22-geo}.
Each DOM houses a photomultiplier tube (PMT) to detect the Cherenkov
light, electronics for pulse digitization and other functions, and
remotely-controllable calibration light sources.  To reduce the impact
of PMT signals due to random noise, only detected signals with minimum 0.25
single photoelectron (p.e.) PMT pulse height were digitized by two
types of waveform digitizers {\it in situ}: the ATWD (Analog Transient
Waveform Digitizer) and an fADC (Fast Analog to Digital Converter).
The time resolution of the ATWD (fADC) is about 3.33~ns (25~ns) with a
readout time window of about 450~ns (6.4~$\mu$s).  Thus the ATWD is
used to capture detailed waveform information on a short time scale
while the fADC records less detailed information on a longer time
scale.  The ATWD also supports three channels with different gains
(x16, x2, and x0.25) to extend its effective dynamic range.

To further remove random noise the digitized signal in a DOM was
required to be in close temporal coincidence with a signal in
neighboring DOMs.  The signals satisfying such a temporal condition in
hardware are called LC (Local Coincidence) hits.  LC hits were then
checked to see whether or not they satisfied a software-based trigger
that selected for patterns potentially produced by a charged particle.
Groups of hits that satisfied a trigger condition were packaged into
``events.''  Higher-level ``filter'' algorithms were applied to each
event, and those events passing one or more filter conditions were
transmitted over satellite to the northern hemisphere for higher-level
analysis.  However, all the data satisfying the software trigger
conditions were stored on tape and shipped to the northern hemisphere.
The software trigger and filter conditions applied to the data used in
this analysis are described in the section below.  For more detail on
the design, construction and performance of IceCube in general,
see~\cite{PMT-paper,DOM-paper,DAQ-paper,HalzenKlein} and references
therein.

%----------------------------   
%     The Data
%---------------------------- 
\section{Data}
\label{sec:data}

\subsection{Experimental Data}
\label{sec:ExperimentalData}

The DOM signals satisfying the LC condition were required by the
online data acquisition (DAQ) system at the surface computing system
in the IceCube Laboratory to satisfy a ``simple majority
trigger'' condition under which eight or more DOMs reported signals in
a 5~$\mu$s time window (``SMT8''). The IC22 trigger rate of 500 to 620
Hz followed the seasonal variation in the cosmic-ray muon flux.  The
DAQ system grouped together DOM hits satisfying the trigger condition
into an event using a broadened $\pm10$~$\mu$s time window.  Triggered
events used in this analysis were accepted if they also satisfied the
extremely high energy (EHE) filter applied to the data online at the
South Pole to reduce low energy events consistent with background.
The EHE filter required $\geq 80$~DOMs registering hits in the event.

We split off about 30\% of the full IC22 dataset (82.4 live-days,
uniformly distributed in time across the data-taking period) to use in
conjunction with simulated data in the design of our subsequent
selection criteria.  In keeping with our procedures for maintaining
blindness in the analysis of data, and thereby reducing human bias in
the analysis of the data, the final result is based on the application
of these selection criteria, unaltered, to the remaining 70\% of the
dataset (200 live-days).

\subsection{Simulated Data}
\label{sec:SimulatedData}

We employed simulated data to develop criteria that enhanced a
possible astrophysical neutrino signal while diminishing backgrounds
from atmospheric neutrinos and cosmic-ray muons.  Exclusive use of
simulated data also permitted us to maintain blindness. For the
signal, the ANIS (All Neutrino Interaction Simulation)
package~\cite{ANIS} was used to produce each neutrino flavor
separately. They were generated following an $E^{-1}$ energy spectrum
to enhance event statistics at higher energy where this analysis is
sensitive.  The neutrinos were propagated through the Earth where 
the Earth shadow effect~\cite{earth-abs} of neutrinos and \nutau 
re-generation~\cite{tau-regen,tau-regen2} were taken into account 
in our simulation.

The events were then run through the IceCube detector
simulation.  The muon (electron) neutrinos were generated over all
zenith angles in the energy range between 10(50)~GeV to 10~EeV while
tau neutrinos were generated between 1~TeV and 1~EeV.

Cosmic-ray muon backgrounds were simulated by generating air shower
events using the CORSIKA package~\cite{CORSIKA}, then propagating the
muons to and through the detector volume with the MMC
package~\cite{MMC}, and finally applying the detector simulation to
the resulting set of particles.

For solitary air showers, a two-component model~\cite{2compo} was
used.  In this model, the entire mass spectrum of cosmic rays is
approximated by only proton and iron components.  Compared to
H\"orandel's polygonato model~\cite{Horandel-2003}, the two-component
model agrees better with experimental data at higher energy (beyond
100 TeV) where this analysis is sensitive.  The cosmic ray primaries
are sampled with an E$^{-2}$ spectrum.  In this way we were able to
produce events more efficiently at the higher primary energies that
contribute most strongly to the background at ultra-high energies.
The cosmic ray flux was then re-weighted to match the expected
spectrum.

The acceptance of IC22 admits the possibility of detecting muons from
multiple quasi-simultaneous air shower events, so we also simulated
muons from two coincident air shower events.  (Higher multiplicities
occur at a negligible rate in IC22 and were not simulated.) For
coincident air showers, H\"orandel's polygonato model of cosmic
rays was used.  Solitary(coincident) atmospheric
air showers were generated with energies between 10(0.6)~TeV--100~EeV and
zenith angles between 0--90$^{\circ}$.

After event generation and detector simulation, the simulated data
were processed in the same way as real data, i.e., with sequential
applications of trigger and filter conditions, as described earlier.

%----------------------------
%    Analysis
%---------------------------- 
\section{Tau Neutrino Identification}
\label{sec:nutauID}

%----------------------------                          
%   Selection Criteria
%----------------------------
\subsection{Selection Criteria}
\label{sec:SelectionCriteria}

\begin{figure}[h!]
\begin{center}
\includegraphics[width=80mm]{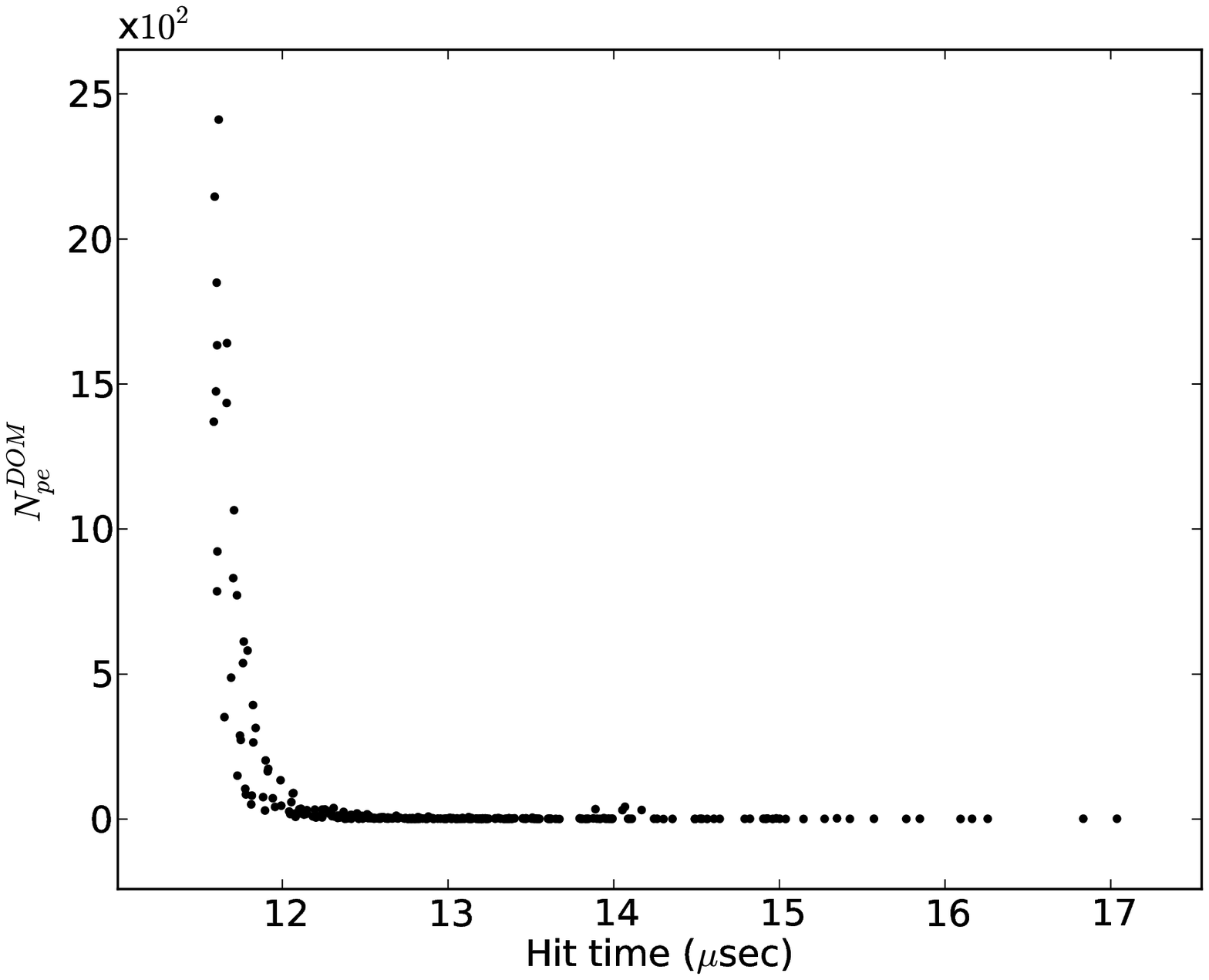} \\
\includegraphics[width=80mm]{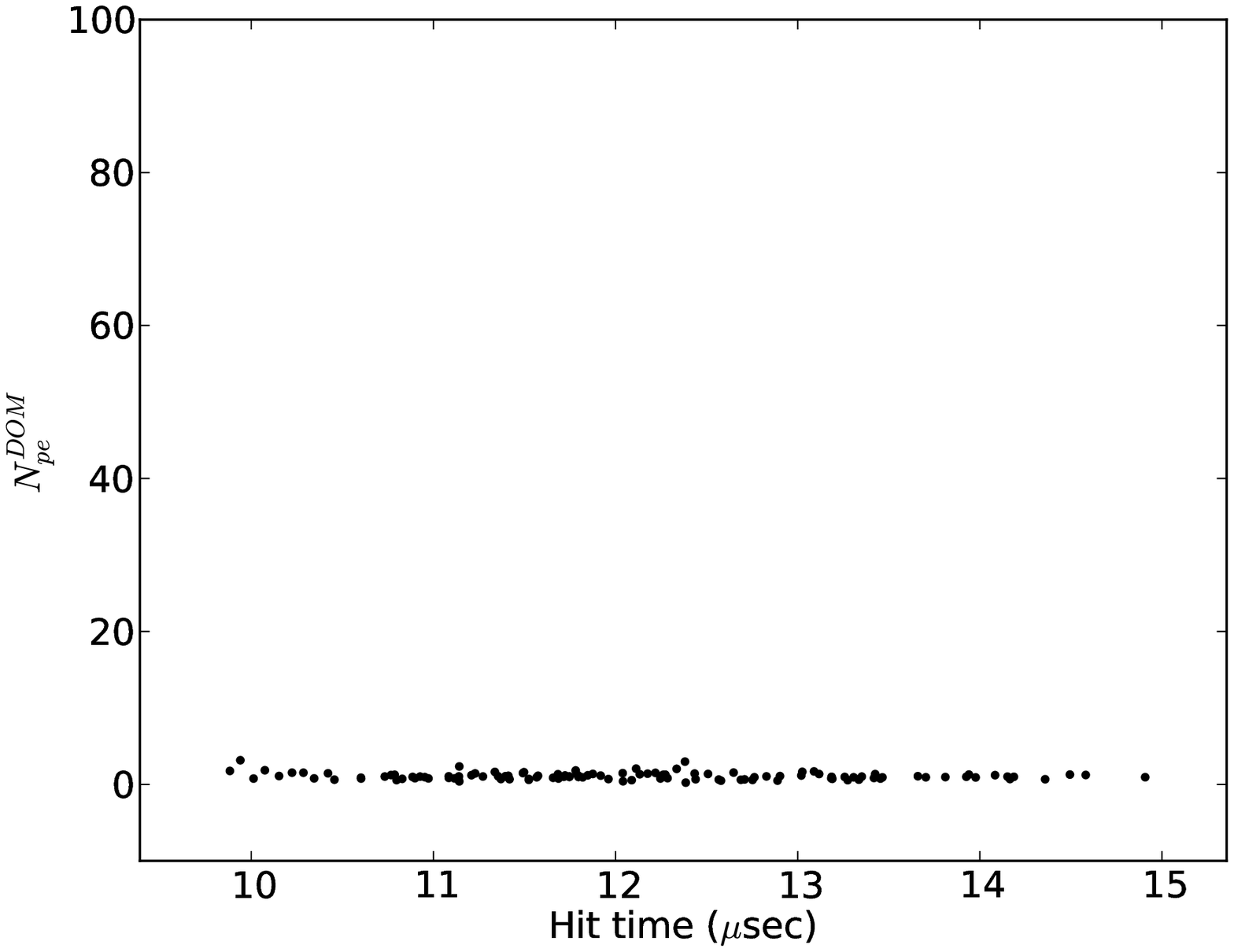}
\end{center}
\caption{\label{fig:char-IRmax-ehe} The quantity \NpeDOMvsT for a
  simulated inverted lollipop (top) and a simulated muon event
  (bottom), with primary particle energies of 25.4~PeV and 2.38~PeV,
  respectively.  The peak of the top plot is at roughly 2500
  photoelectrons.}
\end{figure}

Based on the characteristics of simulated data, we formulated several
event selection criteria to exploit the UHE \nutau signatures of a
track plus one or two showers, in contrast to conventional pure
track-like or pure shower-like events.  Two such criteria use the
reconstructed total number of photoelectrons (\npeNS) per DOM.  The time
associated with such a multi-photoelectron deposit in each DOM is the
time of the first reconstructed photoelectron it detected.  Looking at the full
event time window, \npe for each DOM is plotted vs. time and denoted
\NpeDOMvsTNS.  Fig.~\ref{fig:char-IRmax-ehe} shows \NpeDOMvsT
for a simulated inverted lollipop (top) and a simulated muon event
(bottom).  Note that the times of the hits are with respect to the
event trigger time which has an extended readout time window of $\pm
10$~$\mu$s in IC22.  For this reason, all the hit times exhibit at least
a 10~$\mu$s offset.

\begin{figure}[h!]
\begin{center}
  \subfloat{\includegraphics[width=3.in]{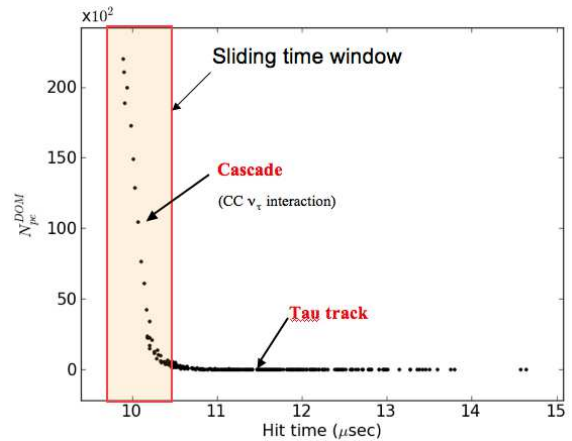}}
\end{center}
\caption{\label{fig:IRmax-ehe} (Color online) 
The maximum current ratio (\irmax) for an event is calculated 
by finding the maximum ratio of charge inside a sliding time window 
to the charge outside this window.  This variable is expected to be 
larger for \nutau events (as in the example shown here) 
than for background events due to atmospheric muons.} 
%  See text for the explanation.}
\end{figure}

\begin{figure}[h!]
\begin{center}
  \subfloat{\includegraphics[width=86mm]{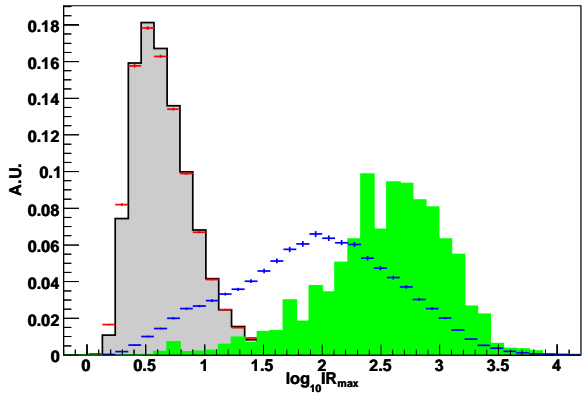}}
\end{center}
\caption{\label{fig:IRmax-sigBG-ehe} (Color online) The logarithm of the \irmax 
  parameter for simulated signal (green histogram for lollipop and blue points for
  all \nutau events) and background (red points for atmospheric muon)
  events, and for data (gray histogram) passing the EHE filter.  
  ÃThe distributions have been normalized to unit integrals to highlight
   the separation between signal and background.  The \irmax
  distributions of inverted lollipop and double bang events are also
  well-separated from the background.}
\end{figure}

To exploit the power of \NpeDOMvsTNS, we devised a parameter called
``maximum current ratio'' (\irmaxNS), defined as the maximum of ${I_{\rm
    in}}/{I_{\rm out}}$ where $I_{\rm in(out)} = {\Delta Q_{\rm
    in(out)}}/{\Delta T_{\rm in(out)}}$.  Here, $\Delta Q_{\rm in}$
was the charge, measured in photoelectrons (p.e.), collected by the DOMs in a
sliding time window of length $\Delta T_{\rm in}$.  The time window
was optimized in this analysis to be 1.2~$\mu$s long.  The
corresponding ``out'' variables were the charge and time measured
outside the sliding time window (see Fig.~\ref{fig:IRmax-ehe}).  As
shown in Fig.~\ref{fig:IRmax-sigBG-ehe}, \irmax is small for
track-like events and large for events containing showers, such as
those produced by \nutauNS.  Since the \irmax cut is related to energy,
it will be applied to data as the last cut together with the other
energy related cut explained at the end of this sub-section.

\begin{figure}[h!]
\includegraphics[width=80mm]{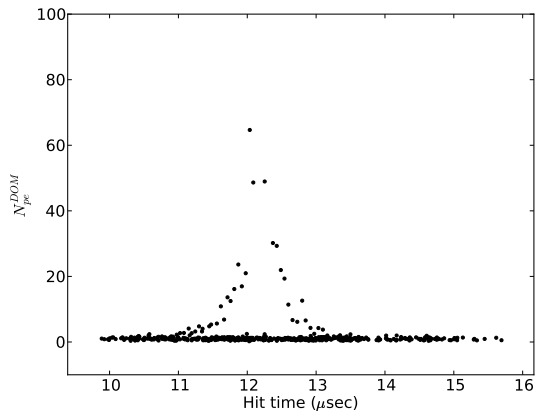}
\caption{\label{fig:brem-muon} The quantity \NpeDOMvsT for a simulated
  muon event, arising from a 336~PeV cosmic-ray primary, with a
  high energy bremsstrahlung energy loss.}
\end{figure}

Although \irmax is very effective at distinguishing most simple
track-like background events from signal events, highly energetic
muons can stochastically deposit large amounts of energy along their
track lengths via bremsstrahlung, pair production, or photonuclear
interactions, potentially mimicking \nutau events.
Fig.~\ref{fig:brem-muon} shows an example of simulated muon with such
a bremsstrahlung whose \irmax value could be similar to that of a
\nutauNS.  
%Seo
Theoretically, \nutau events are most likely to have a large
\NpeDOMvsT at one or both of the temporal edges of the event.  In
practice, \nutau events had a large \NpeDOMvsT in the earliest third
due to the presence of highly scattered photons that extended the
temporal edge of the event to much later times.  We expect future
analyses to be able to devise criteria that reduce the impact of these
scattered photons.  

The ``local charge density'' parameter \lqdenNS,
with units of p.e./ns, was introduced to remove events consistent with
a large energy deposit away from either temporal edge.

\begin{figure}[h!]
\begin{center}
\includegraphics[width=80mm]{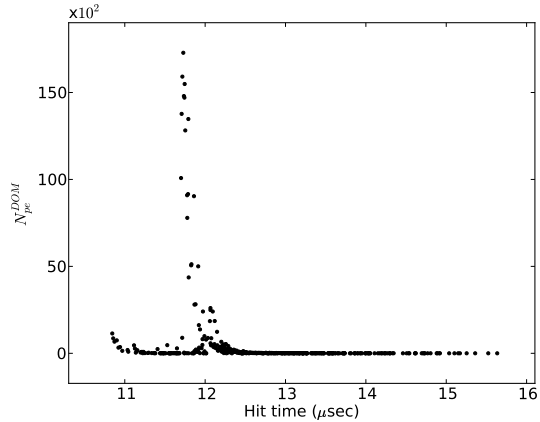}\\
\includegraphics[width=80mm]{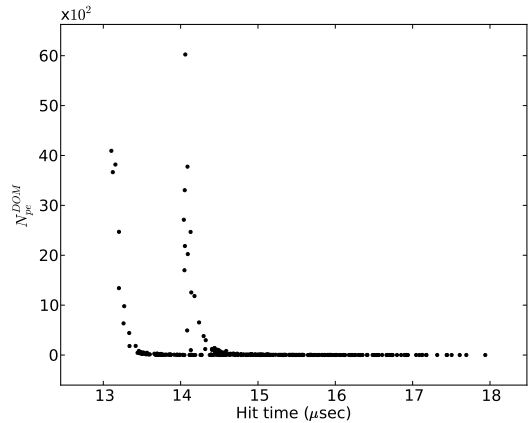}
\end{center}
\caption{\label{fig:localQdenDistribution} The quantity \NpeDOMvsT for
  a simulated lollipop (top) and double bang (bottom) event.  The peak number of
  photoelectrons in the plots above ranges between roughly 6,000 to over 15,000.
  These should be compared to \NpeDOMvsT for a simulated inverted lollipop in
  Fig.~\ref{fig:char-IRmax-ehe} (top) and for an atmospheric muon event in
  Fig.~\ref{fig:brem-muon}.  The atmospheric muon, with a
  bremsstrahlung energy loss roughly in the middle of its contained track length,
  would be rejected by the cut on \lqden described in the text,
  whereas the lollipop, inverted lollipop and double bang would not 
  because the bulk of the detected
  light occurs sufficiently early in the event.}
\end{figure}

\begin{figure}[h!]
\begin{center}
  \subfloat{\includegraphics[width=86mm]{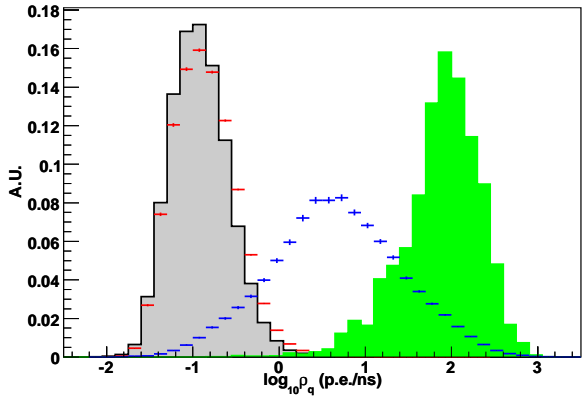}}
\end{center}
\caption{\label{fig:localQdenDistribution-sigBG} (Color online) The logarithm of the
  local charge density parameter (\lqdenNS) for signal (green
  histogram for lollipop and blue points for all \nutau events) and
  background (red histogram for atmospheric muon) events, and for data
  (gray histogram) passing the EHE filter.  The distributions have been 
  normalized to unit integrals to highlight the separation between signal and background.  
  The \lqden distributions of inverted lollipop and double bang events are also 
  well-separated from the background.}
\end{figure}

Partitioning each event into three equal time windows, we calculate
the per-DOM ratios of charge to time in each window.  These ratios are
denoted \lqdenINS, \lqdenII and \lqdenIII in the first, second and
third time window, respectively.  Events for which \lqdenI$\,<5\,$
p.e./ns or \lqdenIII$\,<5\,$ p.e./ns are rejected as being
inconsistent with arising from a \nutau event, since \nutau are
expected to make a significant energy deposition at the beginning
and/or end of its interaction in the instrumented volume.  Events with
small \lqdenII are consistent with arising from \nutau and are not
rejected.  Theoretically, \nutau events are most likely to have a large
\NpeDOMvsT at one or both of the temporal edges of the event.  In
practice, \nutau events had a large \NpeDOMvsT in the earliest third
due to the presence of highly scattered photons that extended the
temporal edge of the event to much later times.  We expect future
analyses to be able to devise criteria that reduce the impact of these
scattered photons.  Figure~\ref{fig:localQdenDistribution} shows \NpeDOM
vs. time and thus illustrates how \lqden can distinguish \nutau events
from muon bremsstrahlung events.
Figure~\ref{fig:localQdenDistribution-sigBG} shows how well \lqden
separates signal from background.

Additional selection criteria were applied to further remove
backgrounds.  The flux of downward-going muons from cosmic-ray air
shower events was reduced by implementing a ``veto layer'' in
software, removing any events in which the average Z position of the
first 4 hits ($\bar{Z}_{\rm init}$), was in the top 50~m of the
detection volume.  Downward-going muons were further removed using the
approximate event velocity $\bar{V}_{Z}$~(m/ns), constructed from the
difference between the positions $Z_{\rm cog}$ and $\bar{Z}_{\rm
  init}$, divided by the difference in their respective times, {\em
  i.e.}, $T_{\rm cog}$ and $\bar{T}_{\rm init}$, where $Z_{\rm cog}$
($T_{\rm cog}$) were the Z position (time) of the center of gravity of
all hit DOMs.  The times here are calculated using the average time of
the hits used to calculate the Z positions.  We removed events
consistent with a downward direction by requiring $\bar{V}_{Z} <
-0.1$~m/ns.

Background events arising from muon stochastic processes at or near the
bottom of the detector, events whose muon tracks may go undetected,
are removed by restricting our sample to events that were reasonably
well-contained in the instrumented detector volume.  We required the
average depth position of all DOMs with signals to satisfy $Z_{\rm
  cog} >-330$~m (as measured from the center of the detector).

We also applied a generic topological selection by calculating the
eigenvalues of the tensor of inertia (ToI) of pulse amplitudes 
(instead of conventional mass)~\cite{TOI} from hit DOMs of each event and keeping
only those events that tended towards sphericity.  Perfectly spherical
events will have three equal ToI eigenvalues, while perfectly
track-like events will have one eigenvalue equal to zero.  We
therefore required that the ratio of smallest eigenvalue to the sum of
all three eigenvalues was $> 0.1$.

\begin{figure*}[h!]
\begin{center}
  \subfloat{\includegraphics[width=60mm]{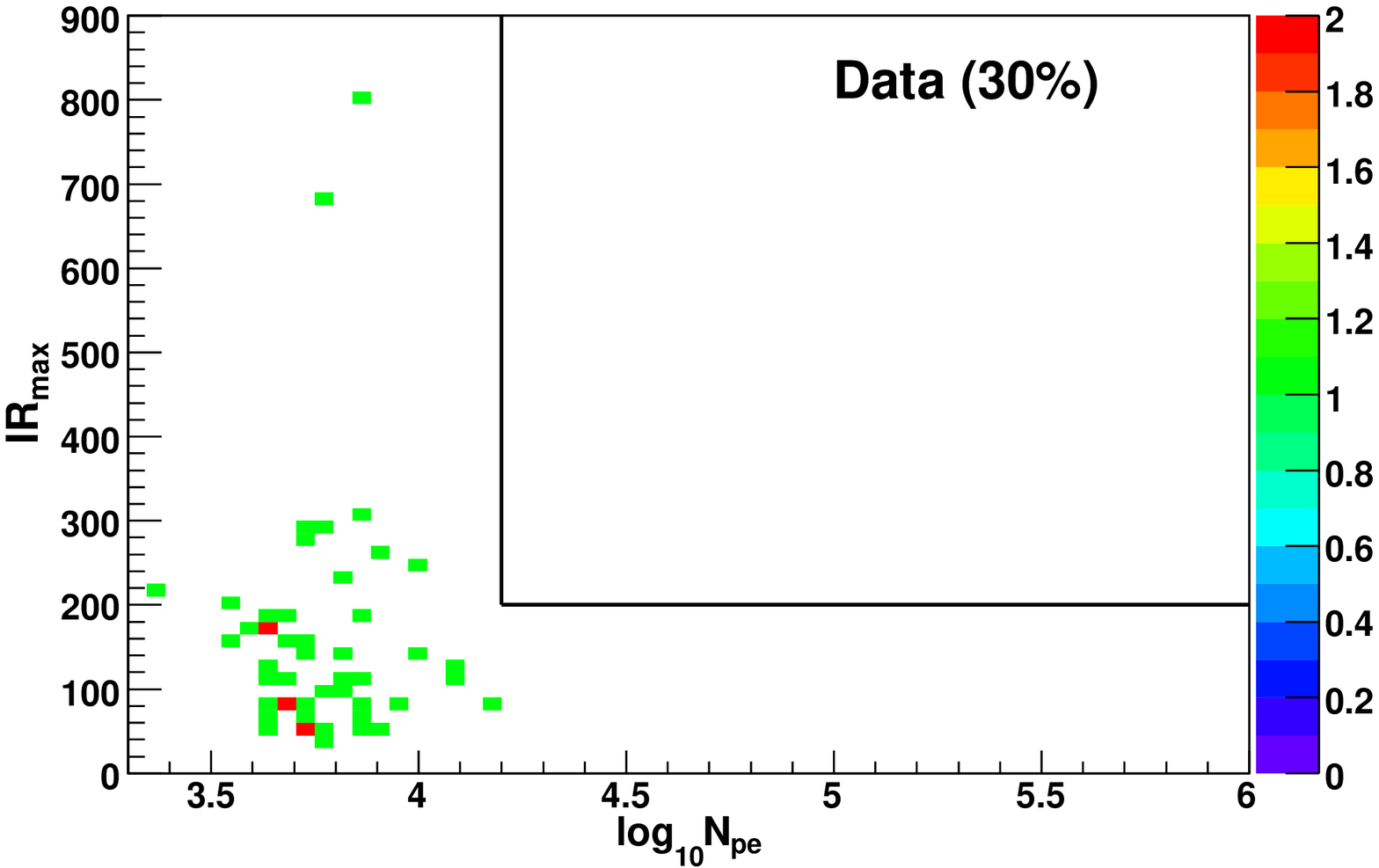}}
  \subfloat{\includegraphics[width=60mm]{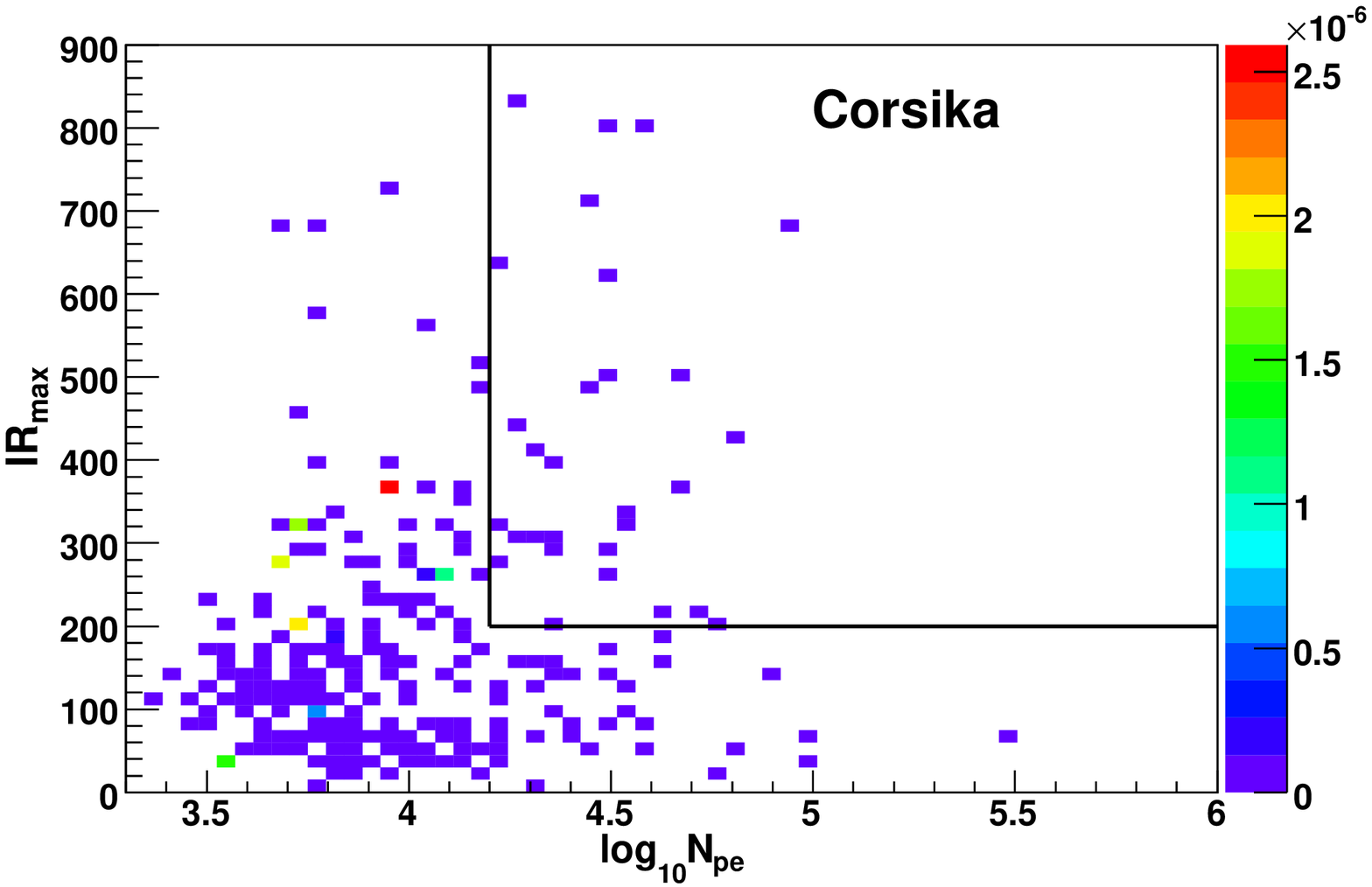}}  \\
  \subfloat{\includegraphics[width=60mm]{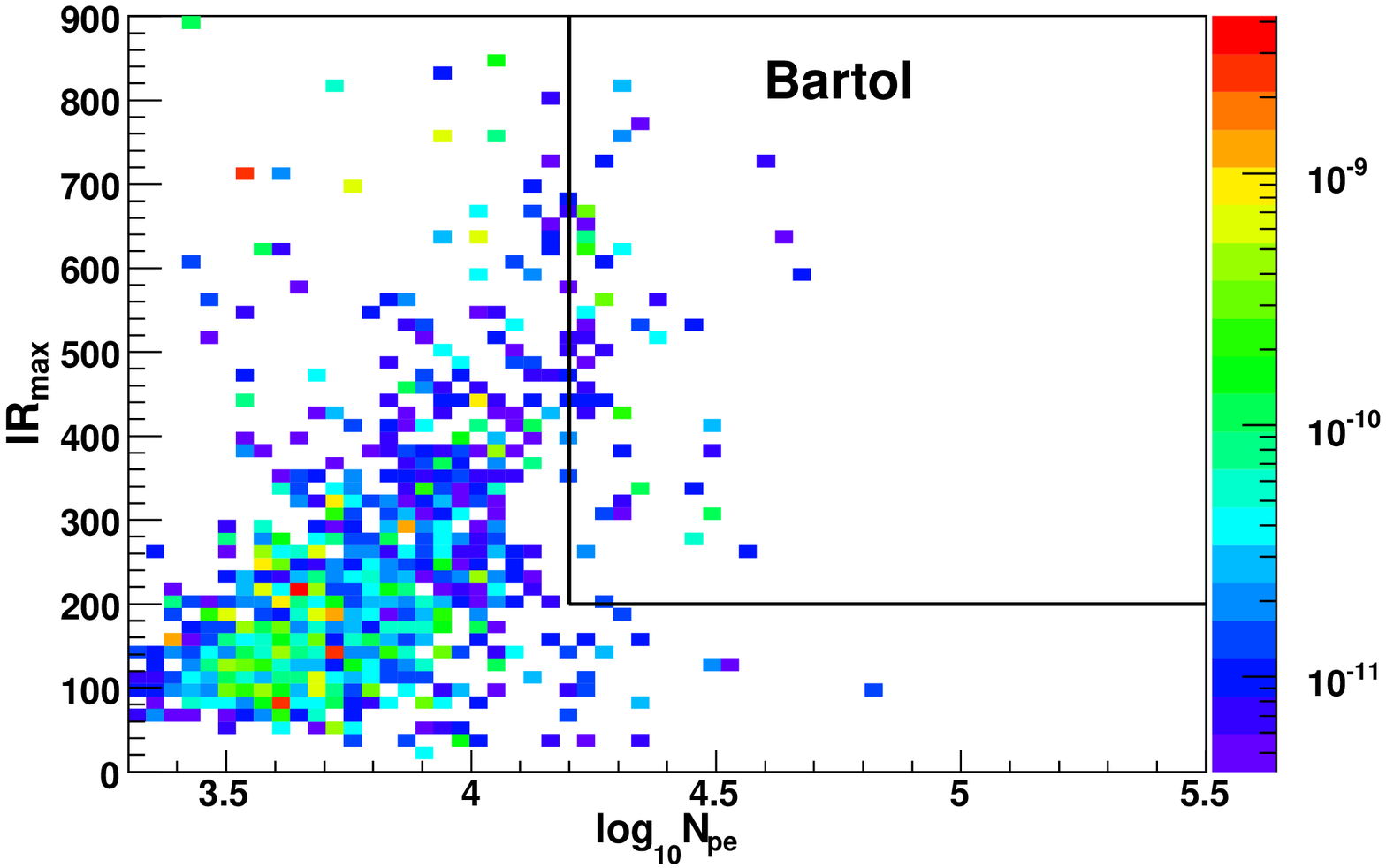}}
  \subfloat{\includegraphics[width=60mm]{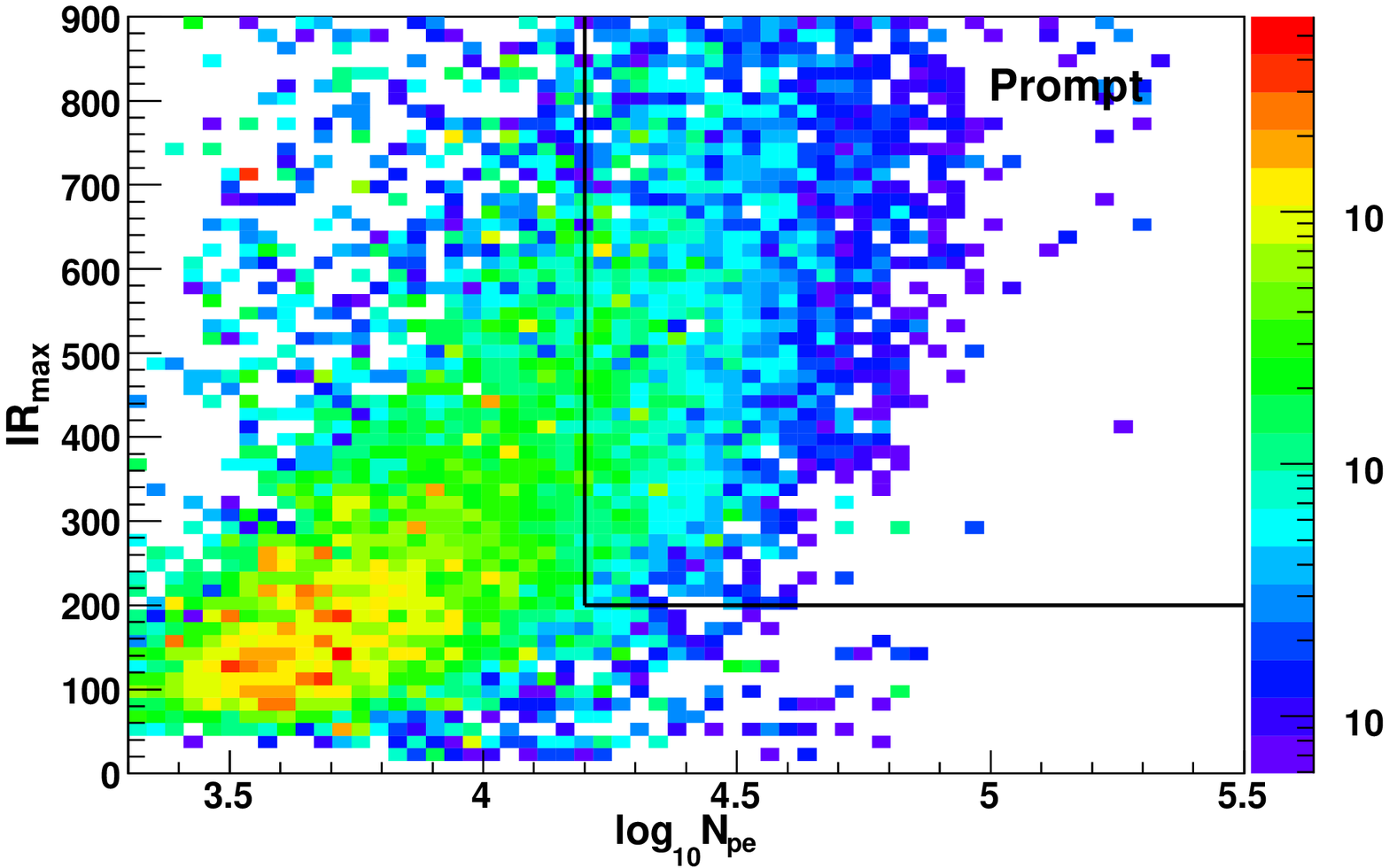}}
  \subfloat{\includegraphics[width=60mm]{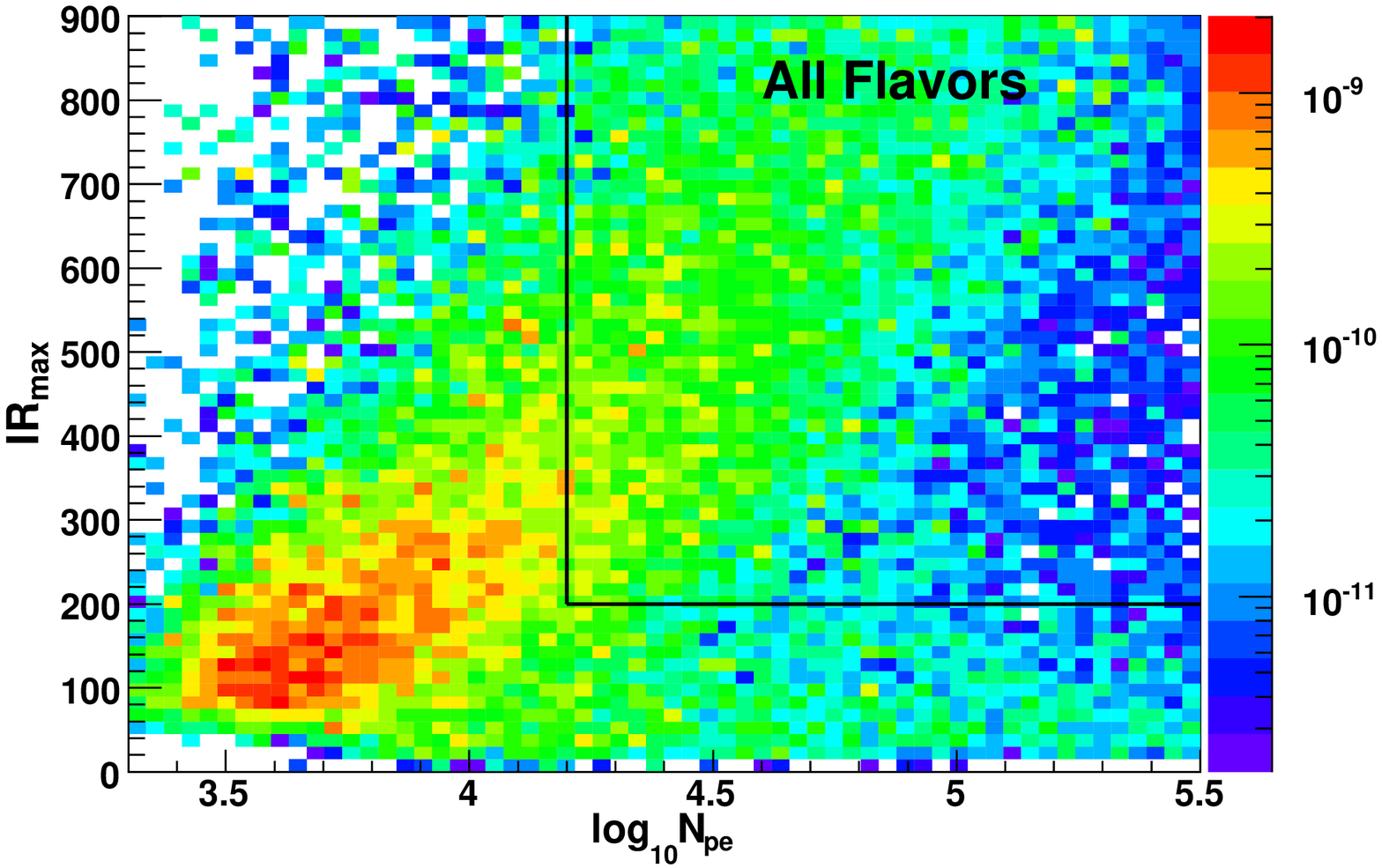}} \\
  \subfloat{\includegraphics[width=60mm]{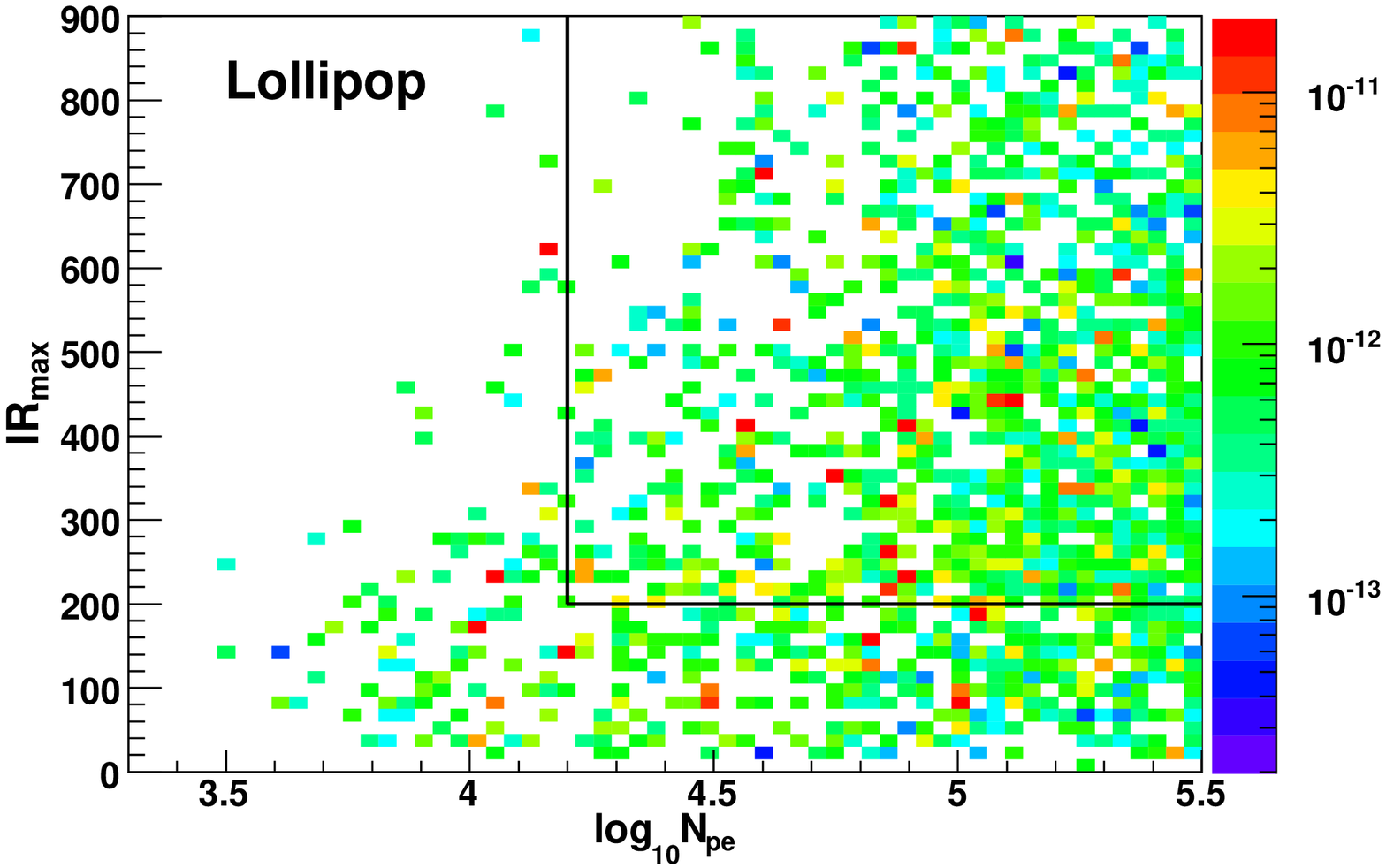}}
  \subfloat{\includegraphics[width=60mm]{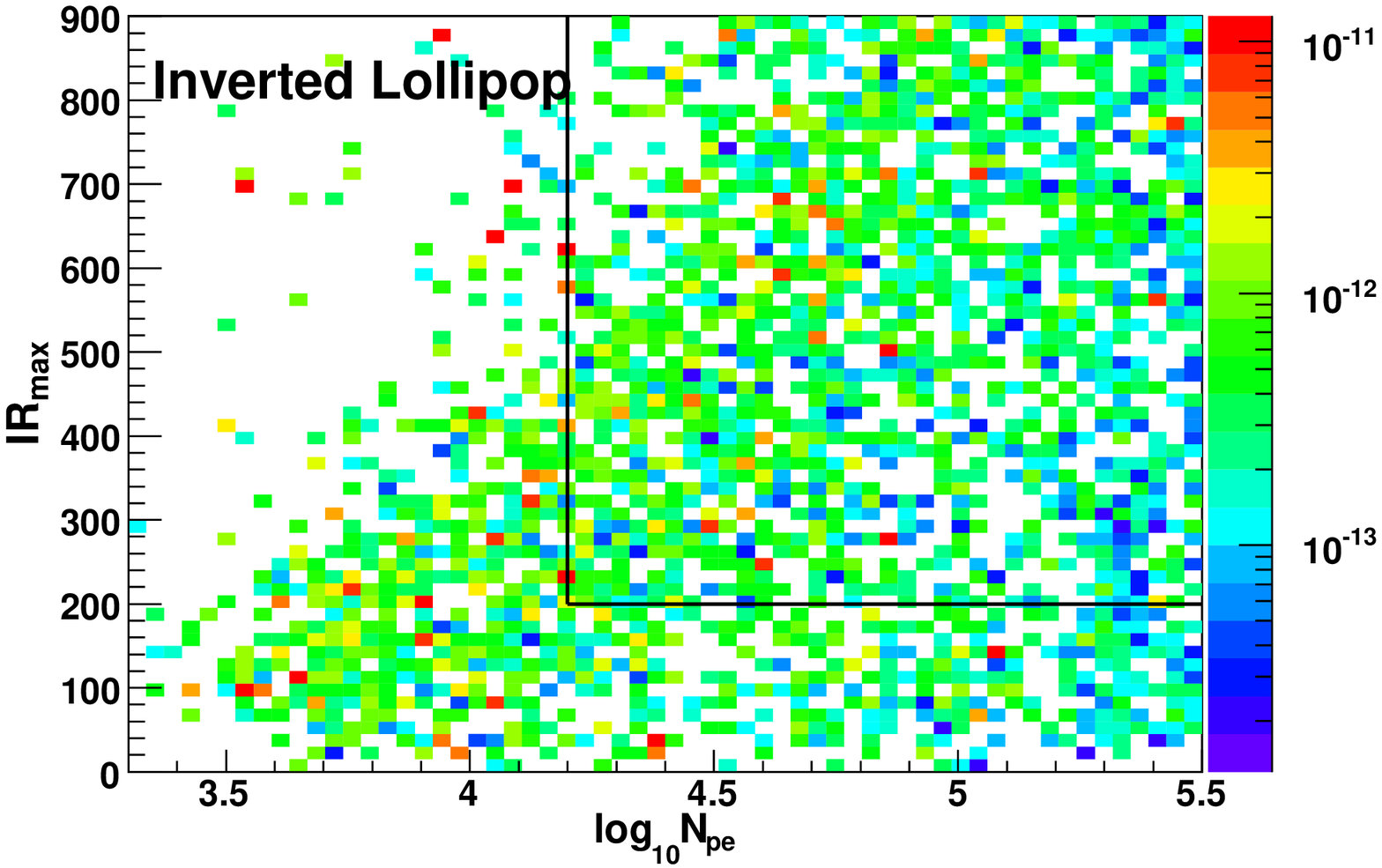}}
  \subfloat{\includegraphics[width=60mm]{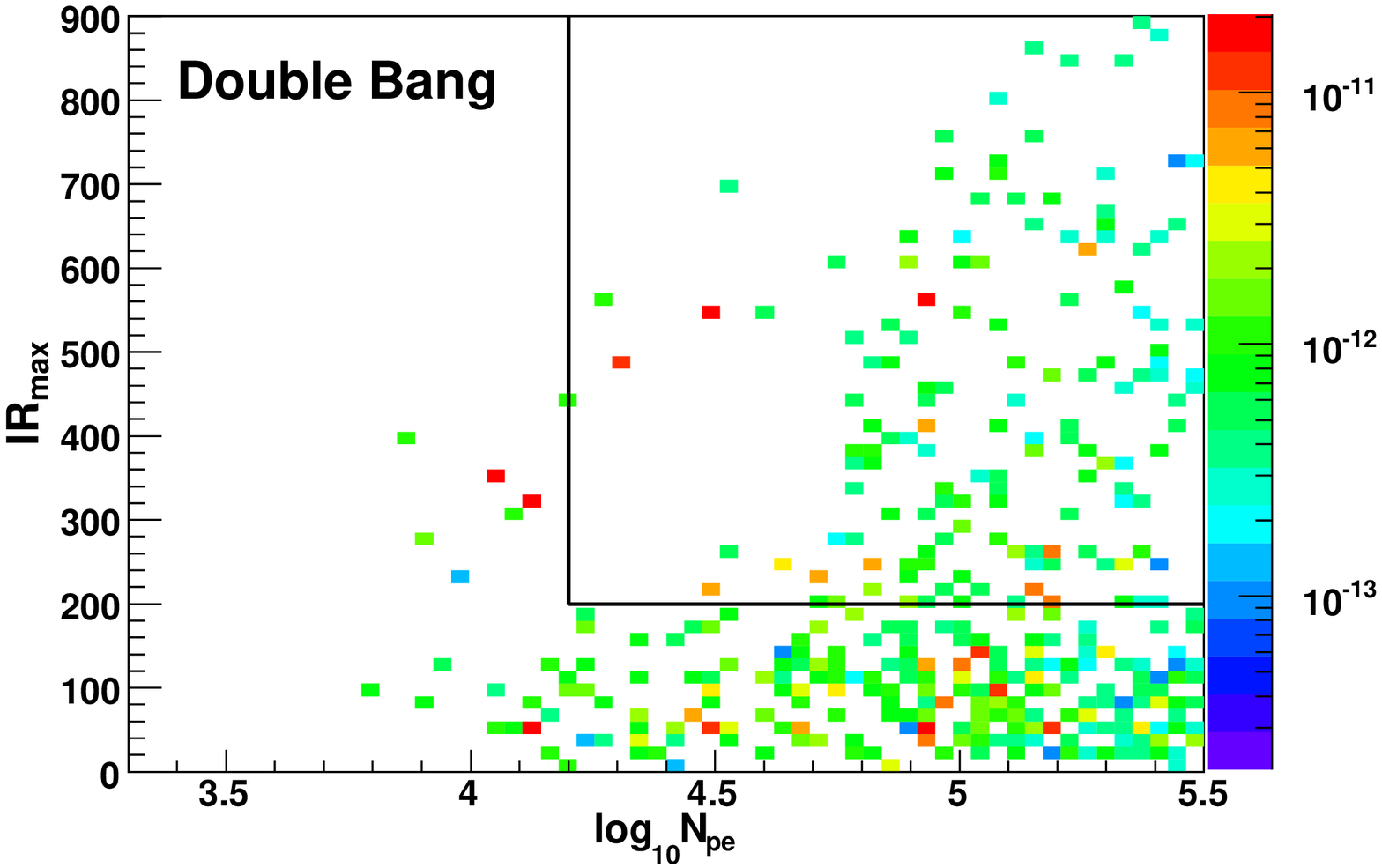}} \\
\end{center}
\caption{\label{fig:dist-beforeFinal}
  (Color online) Distributions of the quantities \irmax vs. \lnpe for 30\% of the
  data (row 1, left), simulated cosmic-ray background (row 1, right),
  simulated atmospheric neutrinos (row 2, left), simulated prompt
  atmospheric neutrinos (row 2, middle), simulated all-flavor neutrino
  signal (row 2, right), and \nutau  lollipop (row 3, left),
  inverted lollipop (row 3, middle) and double bang (row 3, right), assuming an
  $E^{-2}$ spectrum and prior to the final selection criteria optimization.
  The color code represents the event rate in Hz except for the data
  where it represents number of events in 30\% of the data sample
  (82.4 live-days).  The region in the upper right of each plot,
  indicated by the black lines, designates the region selected by the
  optimized criteria as described
  in Section~\ref{sec:OptimizationOfSelectionCriteria}.}
\end{figure*}

Remaining lower energy events were further reduced in number by
requiring a minimum \irmax and \npe for each event.  We required \irmax
$\geq 200$ and \lnpe $\geq 4.2$, the values of which were based on an
optimization that is described in the following section.
Figure~\ref{fig:dist-beforeFinal} shows the distributions of these two
selection criteria for simulated signal, simulated background, and
30\% of the data, prior to the overall optimization of all the
selection criteria.

The selection criteria described above are summarized in
Table~\ref{tab:SummarySelectionCriteria}.

\begin{table*}
   \caption{Summary of the selection criteria used in this analysis.}
   \label{tab:SummarySelectionCriteria}
   \begin{tabular}{|l|l|}\hline
     Selection criterion: & Purpose: \\ \hline\hline
     NDOM $> 80$ & Selects high energy events that produce light in many DOMs. \\ \hline
     \lqdenINS,\lqdenIII$\,>5\,$p.e./ns & Selects events creating light at beginning and/or end of event. \\ \hline
     $\bar{Z}_{\rm init} < 450$~m & Removes events with initial light depositions high in the detector. \\ \hline
     $\bar{V}_{Z} < -0.1$ m/ns  & Removes events consistent with downward trajectories. \\ \hline
     $Z_{\rm cog} >-330$~m & Selects well-contained events and removes cosmic-ray muons interacting \\
                       & near or below the bottom of the detector. \\ \hline
     ToI $> 0.1$ & Favors events with more spherical than track-like topologies. \\ \hline
     \irmax $\geq 200$ & Selects events with large instantaneous light depositions.  \\ \hline
     \lnpe $\geq 4.2$ & Selects high energy events that produce a large amount of light. \\ \hline
   \end{tabular}
\end{table*}

%\clearpage

%--------------------------------
%     Final Cut Tuning
%-------------------------------- 
\subsection{Optimization of Selection Criteria}
\label{sec:OptimizationOfSelectionCriteria}

\begin{figure}[h]
\includegraphics[width=86mm]{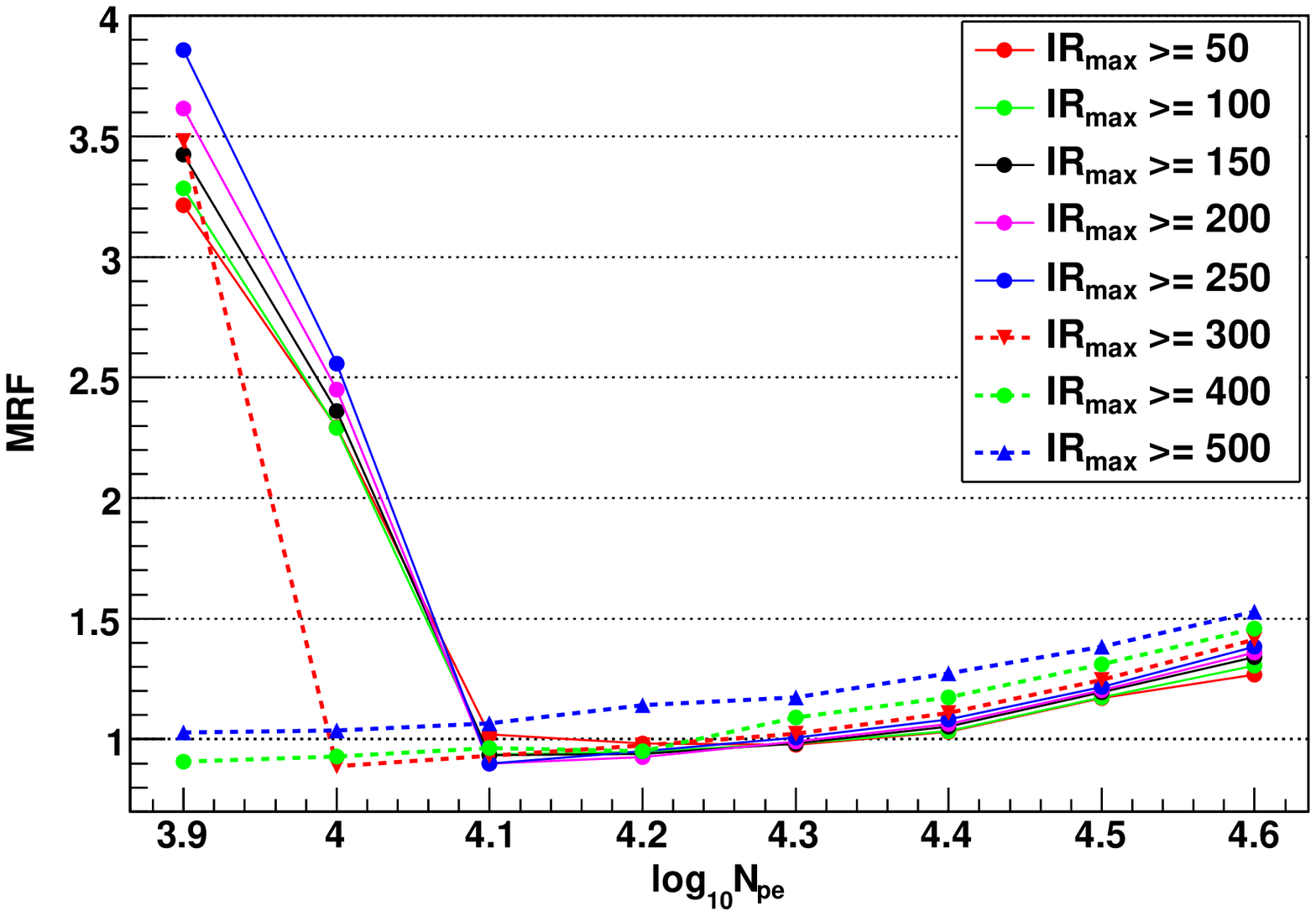}
\caption{\label{fig:tuning} (Color online) We optimized the selection criteria for 
  \irmax and \lnpe using the Model Rejection Factor (MRF) formalism.  The plot
  shows how the MRF varies as a function of \lnpe for different values
  of \irmaxNS.  We chose values for these parameters near but not
  exactly at the minimum shown for reasons explained in the text.}
\end{figure}

The final values for \irmax and \npe were optimized by minimizing the
Model Rejection Factor (MRF)~\cite{MRF} before applying them to the
full dataset.  We varied the values of \irmax and \lnpe as shown in
Fig.~\ref{fig:tuning}, finding a shallow minimum at MRF$\sim 0.89$.
At this MRF, the expected all-flavor signal and background were 3.52
and 0.81 events, respectively, using the Waxman-Bahcall upper bound for signal,
translated to account for what would be detected following standard
neutrino oscillations, of $E^2_\nu \phi_\nu < \frac{3}{2} \times
4.5\times 10^{-8}$~GeV/cm$^2$s~sr~\cite{WB} for the signal neutrino
flux normalization with E$^{-2}$ spectrum.  Assuming standard neutrino
oscillations, we expect one-third of this flux to be \nutauNS. The
corresponding optimized values are \irmax$\geq 300$ and \lnpe$\geq
4.0$.  However, in order to be conservative in the face of limited
simulated event statistics, we chose instead to use \irmax$\geq 200$
and \lnpe$\geq 4.2$, resulting in an MRF$= 0.93$ and expected all
flavor signal and background event counts of 3.18 and 0.60,
respectively.

%-------------------------------- 
%      Cut efficiency
%-------------------------------- 
\subsection{Signal Selection Efficiency}
\label{sec:SignalSelectionEfficiency}

The event rates for the selection criteria described in
Section~\ref{sec:SelectionCriteria} were grouped into sets (EHE, S1-4)
for reference purposes and are summarized in Table~\ref{tab:rate-sig}
for simulated signals.  It is evident from Table~\ref{tab:rate-sig}
that this analysis, though designed to be sensitive primarily to UHE
\nutau signals, also had appreciable sensitivity to UHE \nue and \numu
signals.  The final limit described below will therefore be applicable
to all neutrino flavors.
Figures~\ref{fig:ehe-sigBG}-\ref{fig:set3-sigBG} show the distribution
of event rates (Hz) for each cut parameter for simulated signal as
well as background, and a sample of IC22 data.  All plots show data
after application of the EHE filter (Fig.~\ref{fig:ehe-sigBG}) and
sets of selection criteria S1 (Fig.~\ref{fig:set1-sigBG}), S2
(Fig.~\ref{fig:set2-sigBG}), and S3 (Fig.~\ref{fig:set3-sigBG}).

The efficiency of the event selection criteria for accepting signal
can be obtained from Fig.~\ref{fig:efficiency} (top).  The bottom plot
of that same figure shows the effective area \Aeff for each neutrino
flavor after application of the SMT8 trigger condition and the full
suite of selection criteria.  Using simulated signal, \Aeff is defined
by $\Phi_\nu {\rm A}_{\rm eff} T = {\rm N}_{\rm det}$, where
$\Phi_\nu$ is the neutrino flux prior to any propagation or
interaction effects in the Earth, $T$ is a length of time, and ${\rm
  N}_{\rm det}$ is the number of detected events.  The \Aeff is not
used in the calculation of our limit on UHE neutrino production, but
event rates for a particular theoretical model subject to the
selection criteria in this analysis may be estimated via the product
of the effective area and the model's predicted flux.  In the energy
range pertinent to this analysis, signal events must be either
downward-going or horizontal due to Earth absorption of upward-going
neutrinos for $E_\nu > \sim 100$~TeV.

\subsection{Background Selection Efficiency}
\label{sec:BackgroundSelectionEfficiency}

\begin{figure}[h!]
\begin{center}
  \subfloat{\includegraphics[height=1.6in]{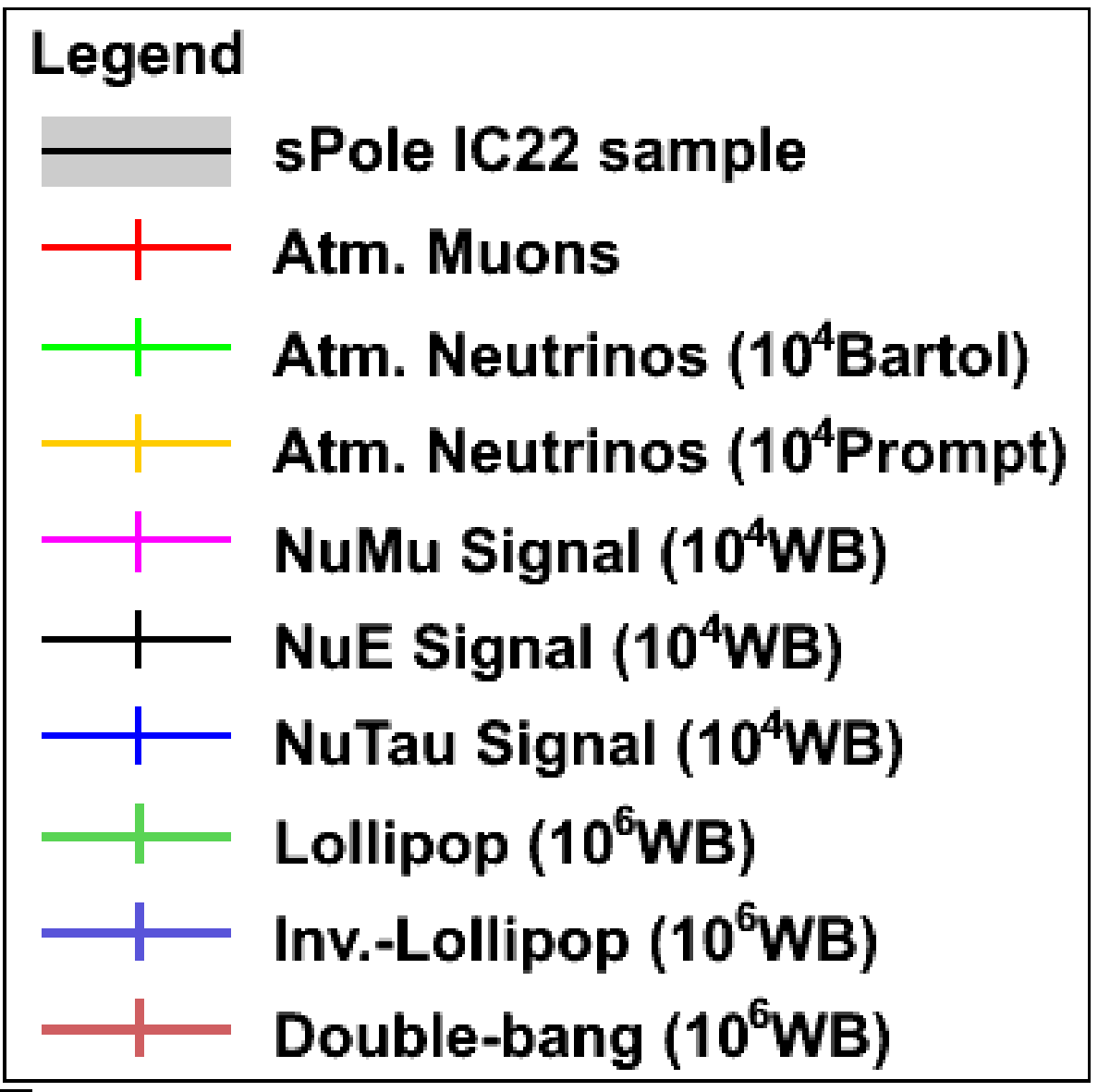}}\\
  \subfloat{\includegraphics[width=86mm]{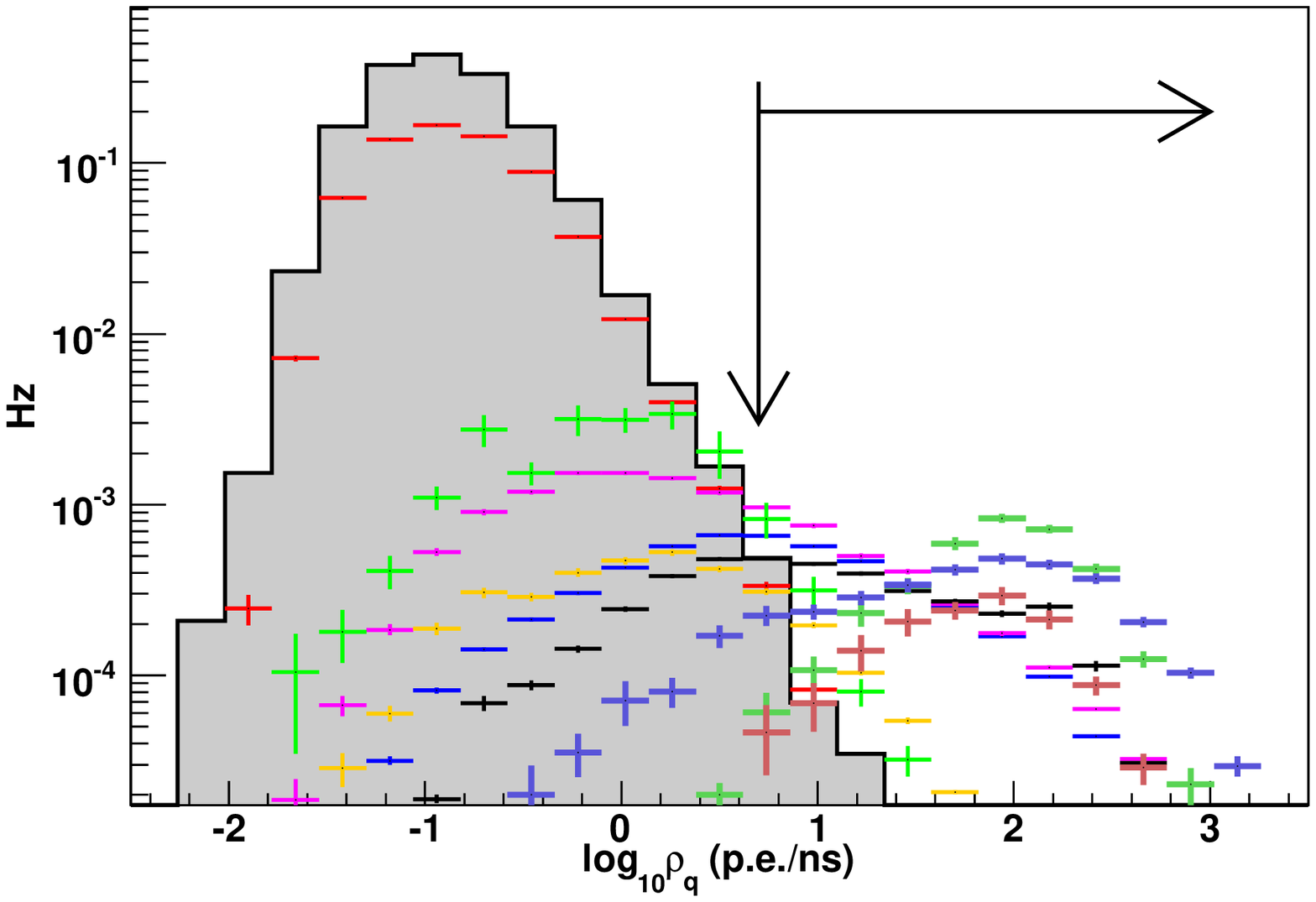}}
\end{center}
\caption{\label{fig:ehe-sigBG} (Color online) Distributions of local charge density for
  signal, BG, and 30\% of the IC22 data just before application of the
  ``S1'' set of selection criteria.  The vertical (horizontal) arrow
  line represents the cut value (selected region) of that set. }
\end{figure}

\begin{figure}[h!]
\begin{center}
  \subfloat{\includegraphics[height=1.6in]{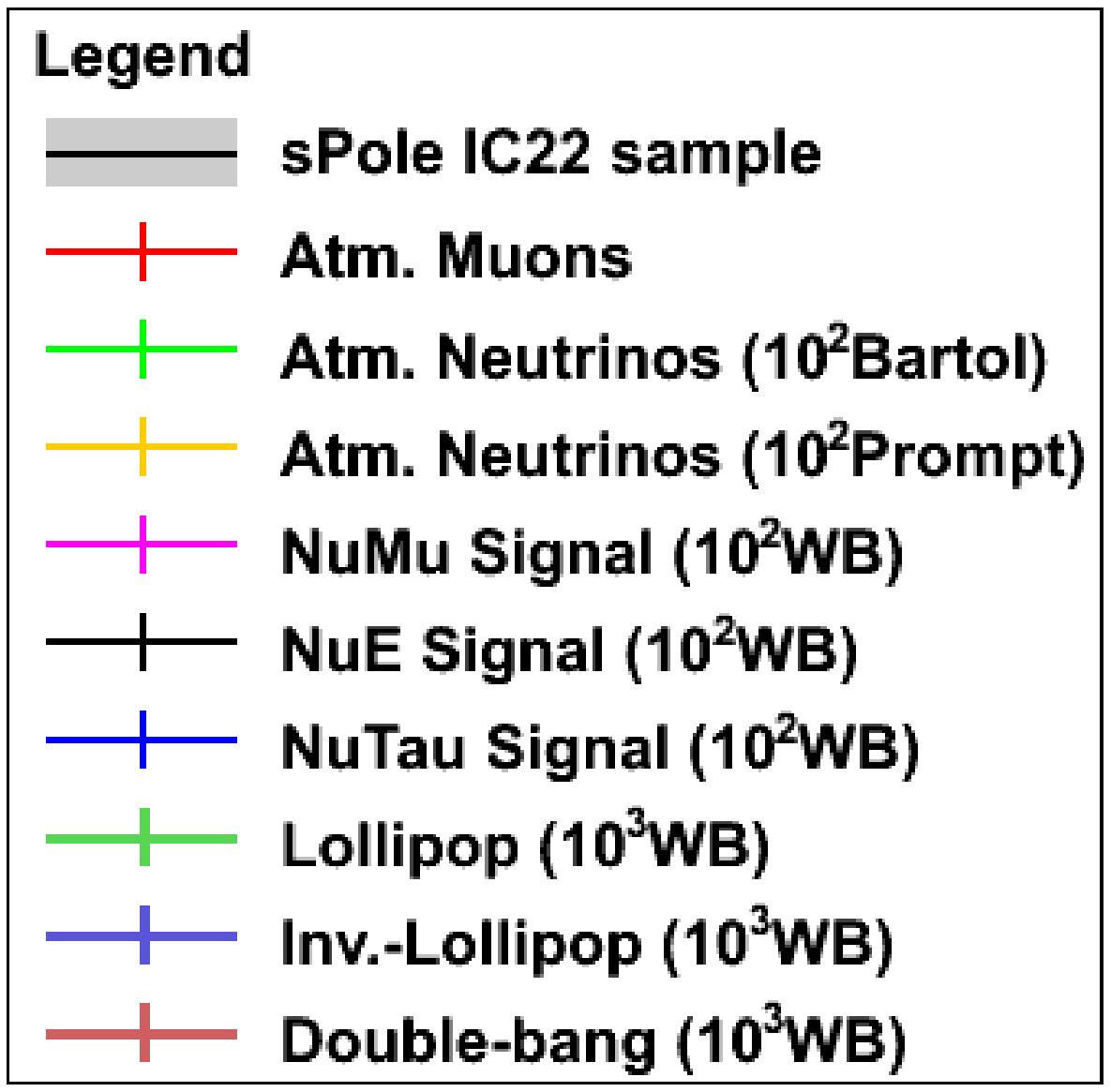}} \\
  \subfloat{\includegraphics[width=86mm]{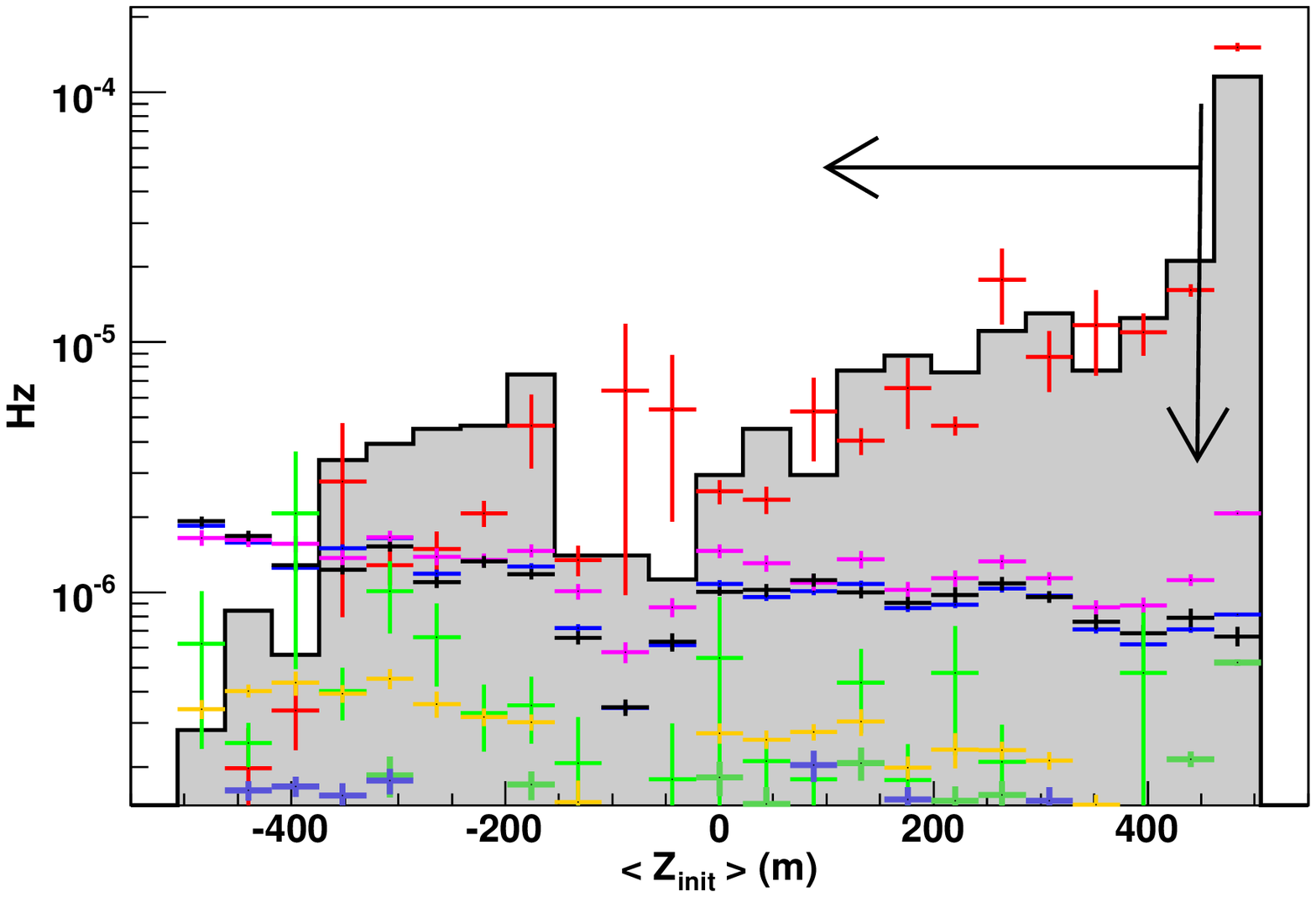}}\\
  \subfloat{\includegraphics[width=86mm]{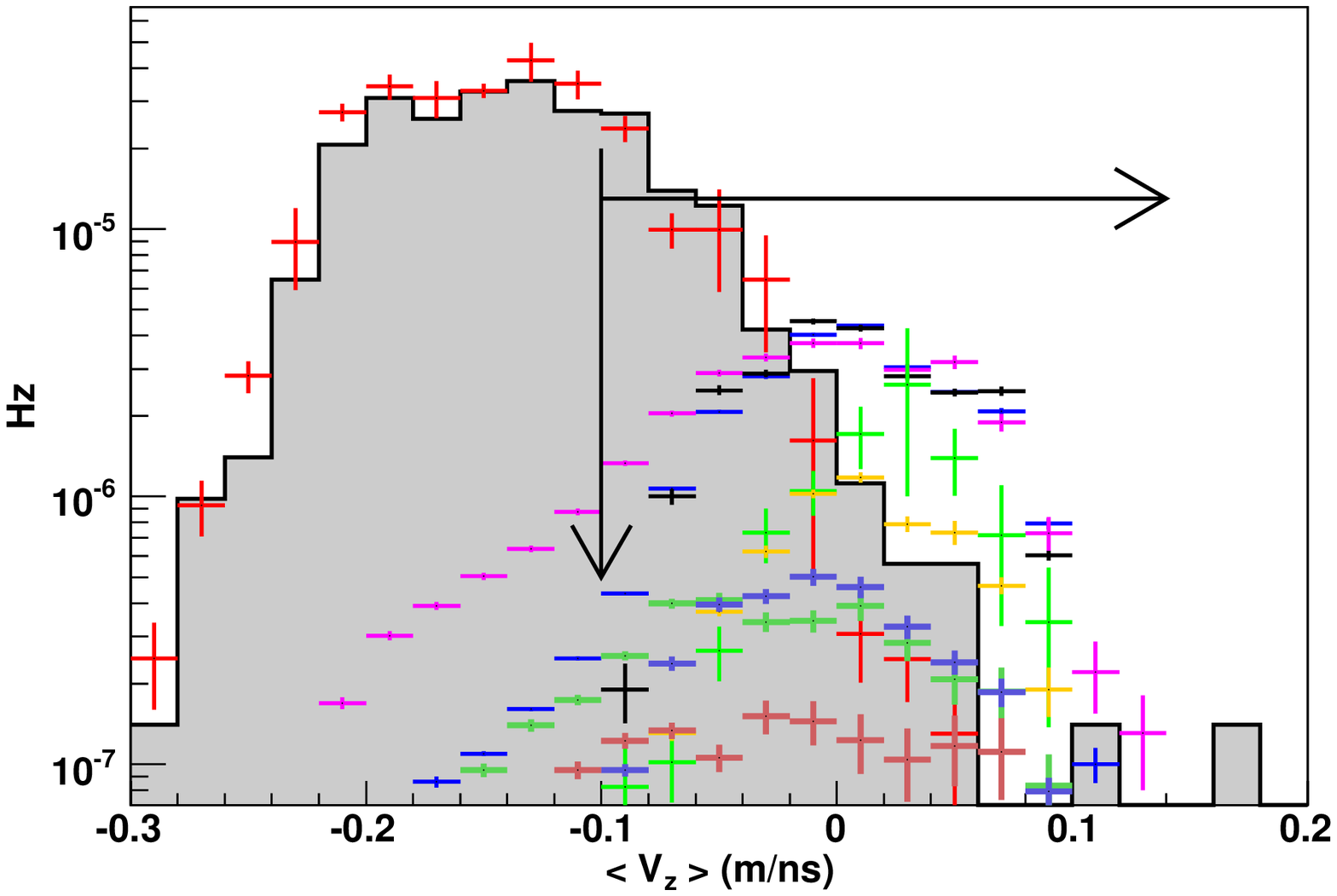}}
\end{center}
\caption{\label{fig:set1-sigBG} (Color online) Distributions of
  the average z-position of initial hits (top) and the average velocity
  z-component (bottom) for signal, BG, and 30\% of the IC22 data just
  before application of the ``S2'' set of selection criteria.  The
  vertical (horizontal) arrow lines represent the cut values (selected
  regions) of that set.  }
\end{figure}

\begin{figure}[h!]
\begin{center}
  \subfloat{\includegraphics[height=1.6in]{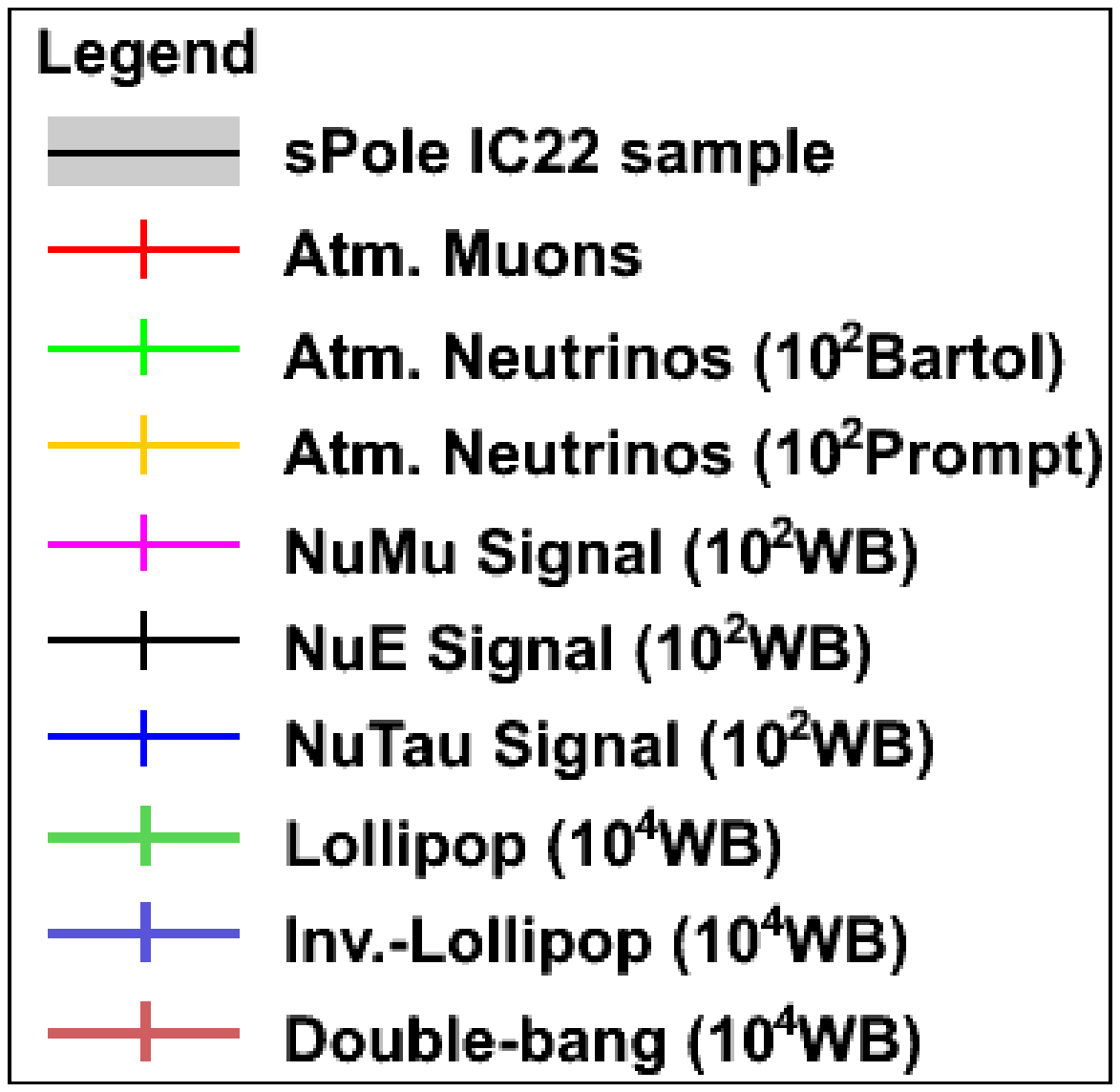}}\\
  \subfloat{\includegraphics[width=86mm]{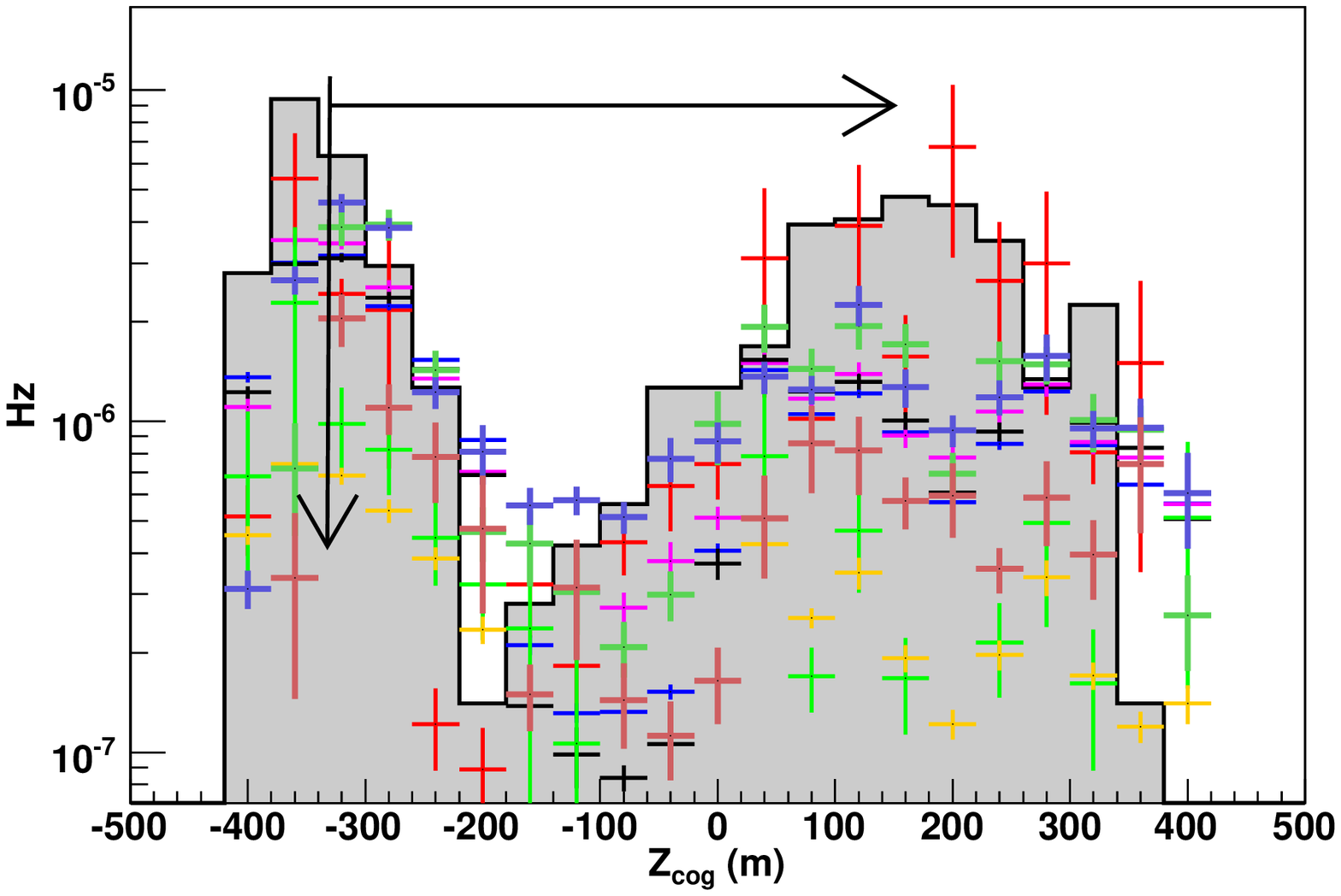}}\\
  \subfloat{\includegraphics[width=86mm]{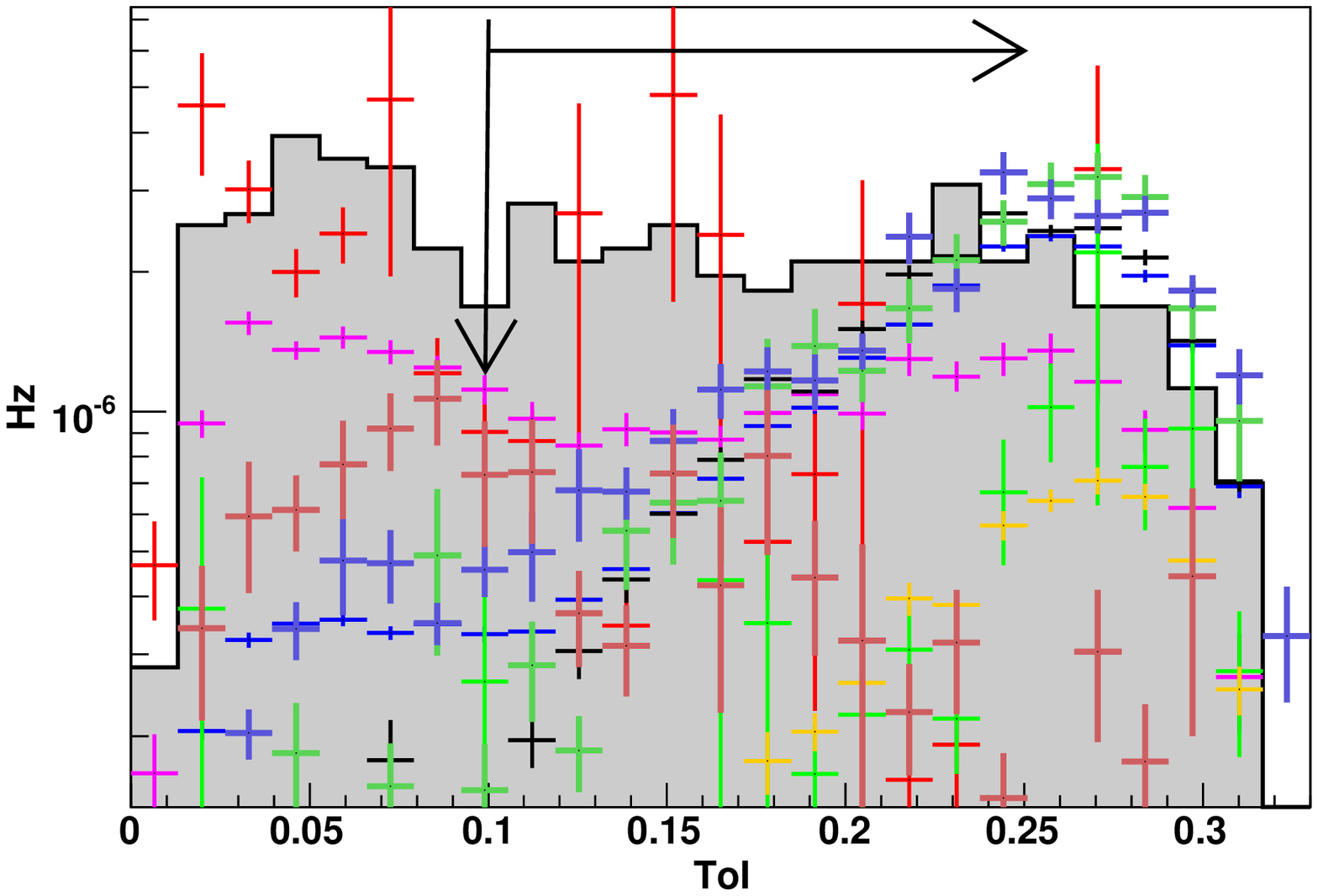}}
\end{center}
\caption{\label{fig:set2-sigBG} (Color online) Distributions of the center
  of gravity of the z-position (top) and tensor of inertia (bottom) for
  signal, BG, and 30\% of the IC22 data just before application of the
  ``S3'' set of selection criteria.  The vertical (horizontal) arrow
  lines represent the cut values (selected regions) of that set.  }
\end{figure}

\begin{figure}[h!]
\begin{center}
  \subfloat{\includegraphics[width=86mm]{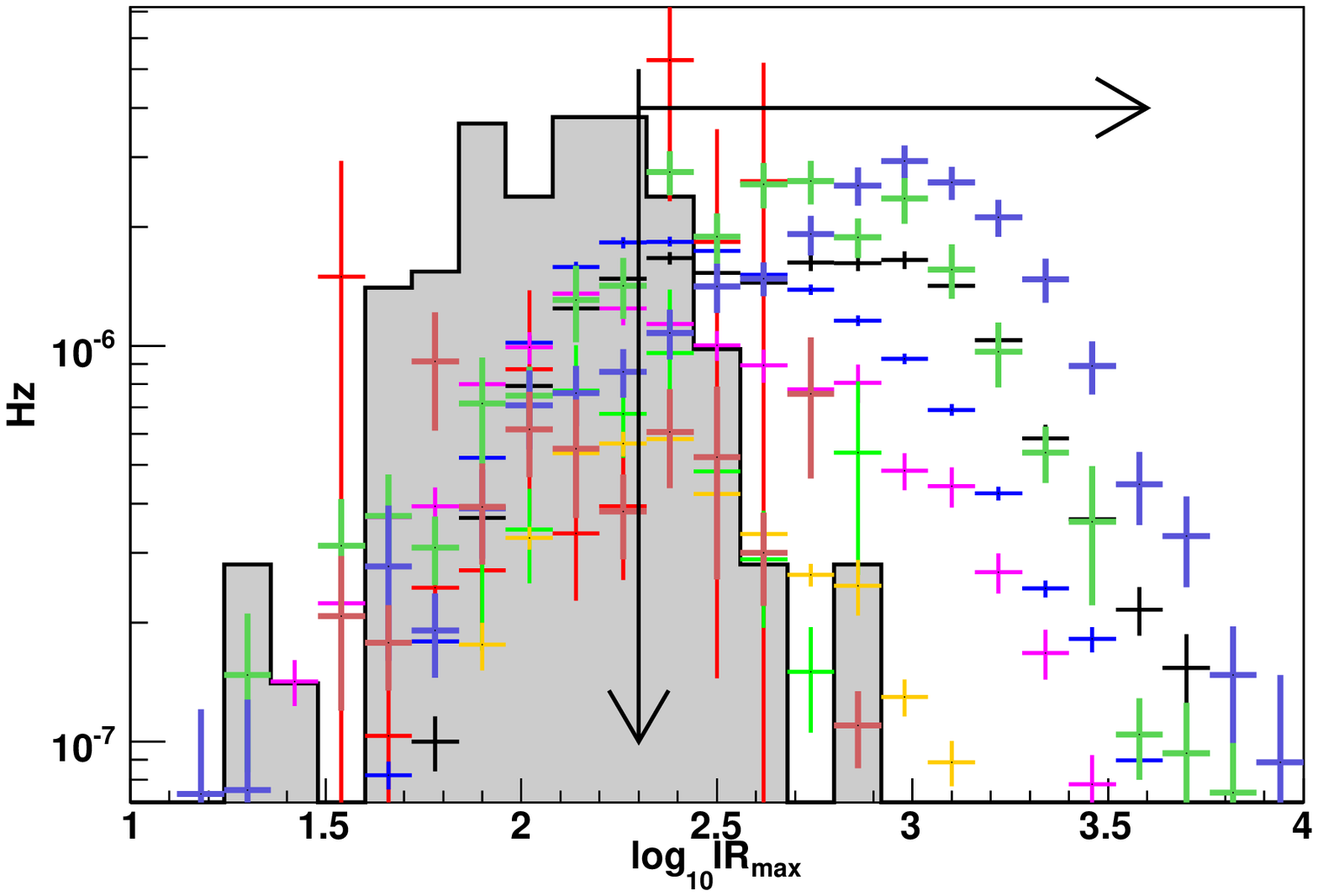}}\\
  \subfloat{\includegraphics[width=86mm]{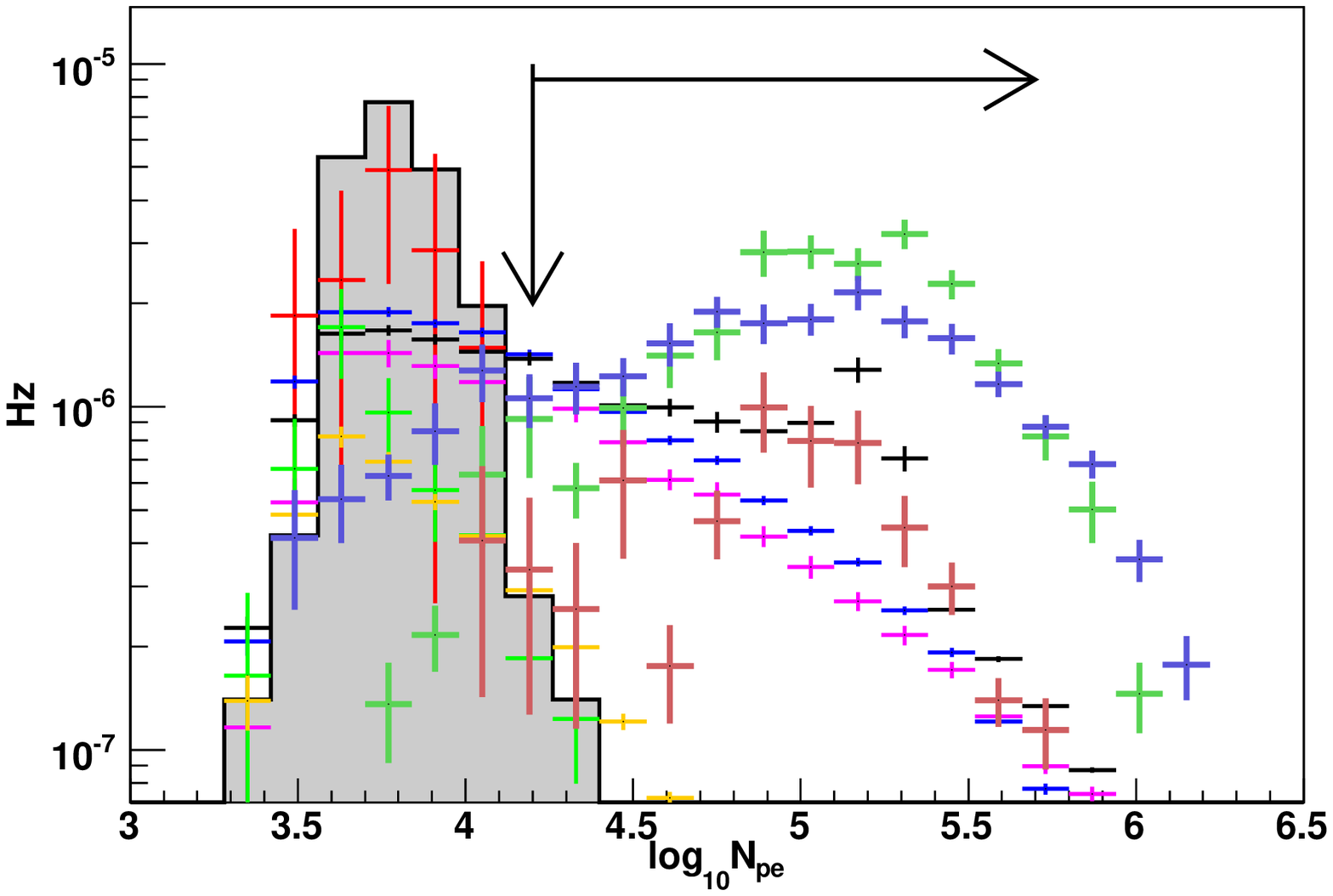}}
\end{center}
\caption{\label{fig:set3-sigBG} (Color online) Distributions of
  the maximum current ratio (top) and the number of photo-electrons (bottom)
  for signal, BG, and 30\% of the IC22 data just before application of
  the ``S4'' set of selection criteria.  The vertical (horizontal)
  arrow lines represent the cut values (selected regions) of that set.
  The legend for these plots is the same as in
  Fig.~\ref{fig:set2-sigBG}.  }
\end{figure}

\begin{figure}[h!]
\includegraphics[width=86mm]{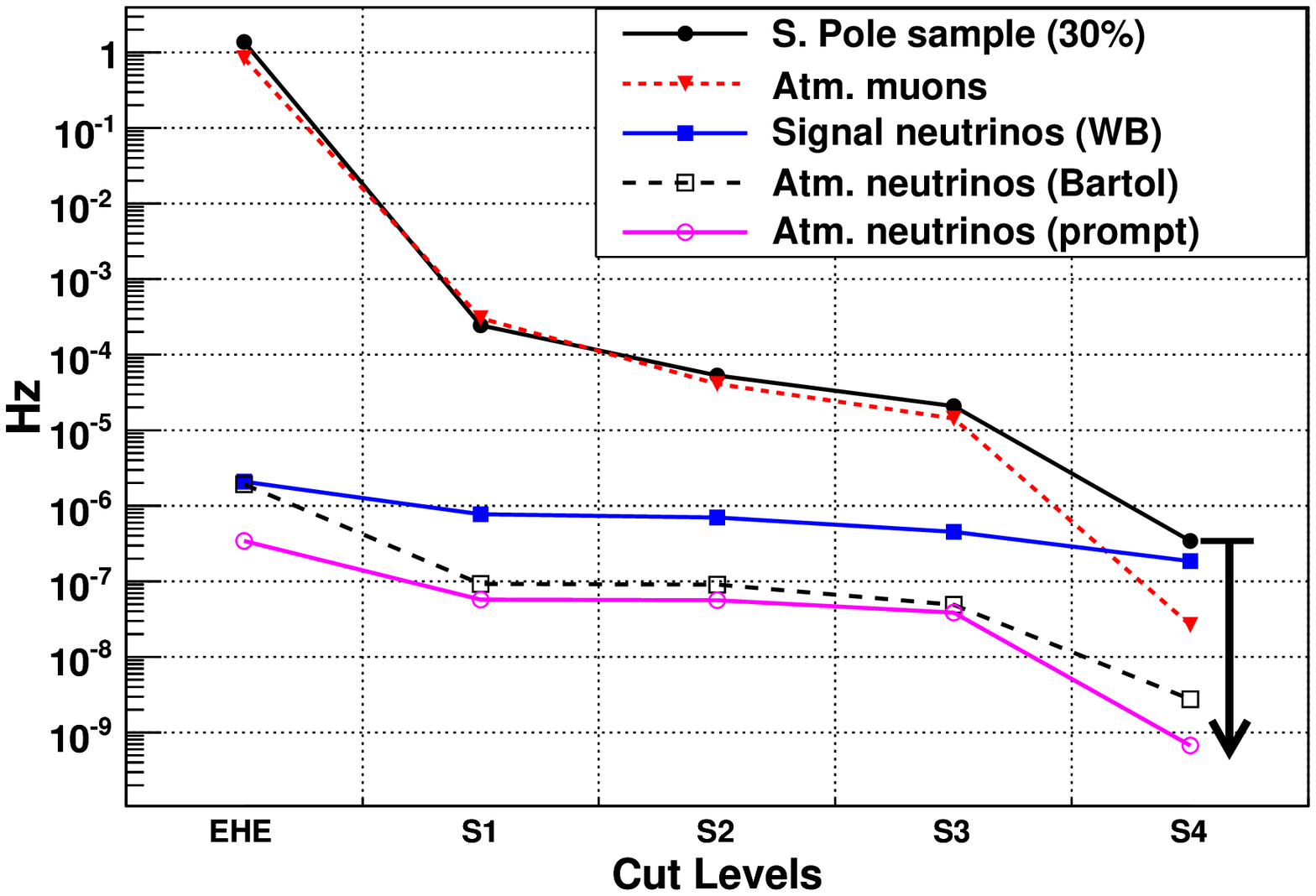}\\
\includegraphics[width=86mm]{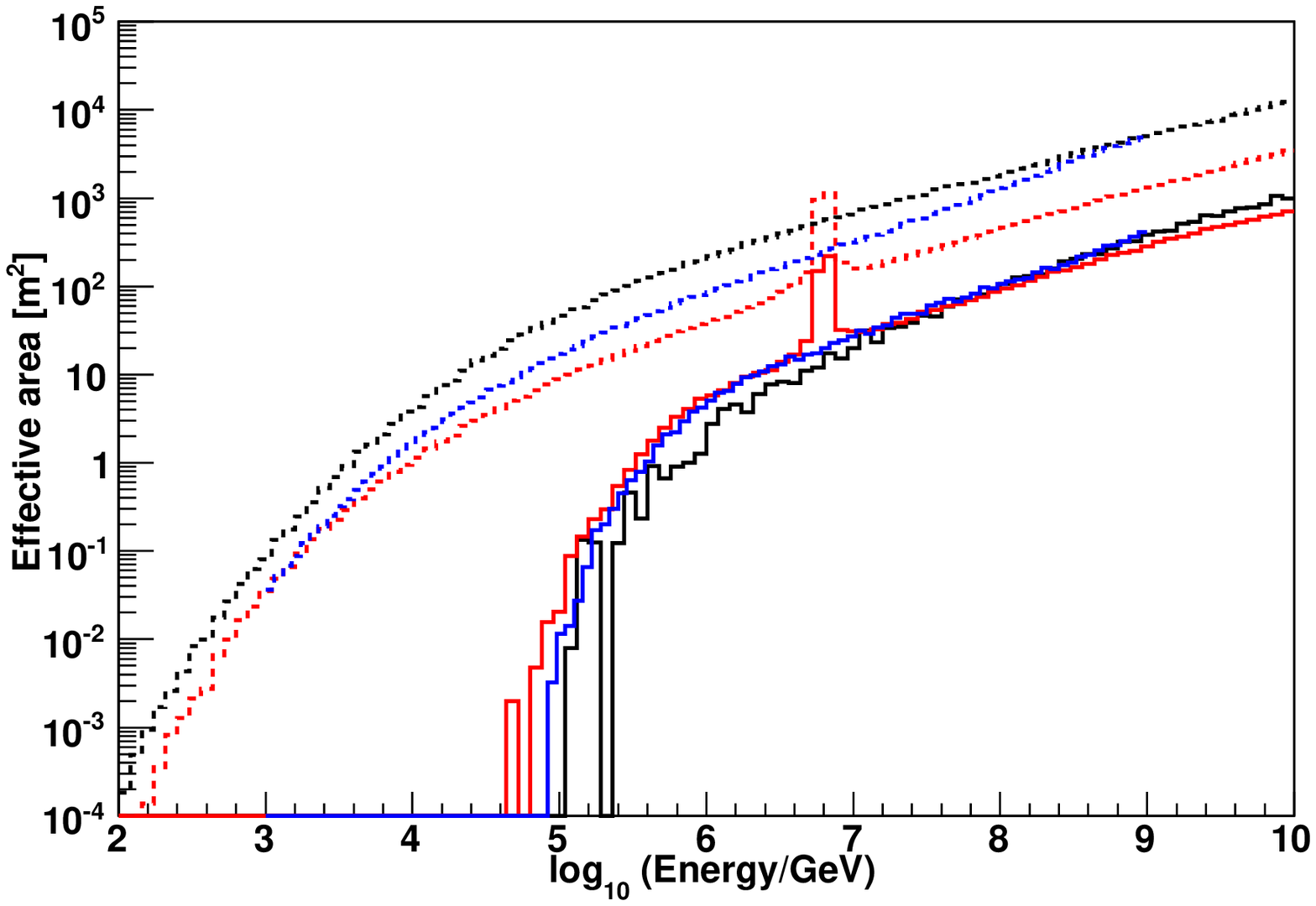}
\caption{\label{fig:efficiency} (Color online) Top: Event rate (Hz) at
  each cut level for simulated signal and background, and 30\% of the
  data sample.  At S4, there were zero events in the data sample, and
  so the 90\%~CL upper limit value was plotted as indicated by the
  black arrow.  Bottom: IC22 effective areas vs. neutrino energy for
  each neutrino flavor (red: \nueNS, blue: \nutauNS, black: \numuNS)
  after application of the SMT8 trigger (dashed lines) and after
  application of all selection criteria (solid lines).}
\end{figure}

The event rates for simulated background and 30\% of the data sample
are summarized in Table~\ref{tab:rate-bg}.
Figures~\ref{fig:ehe-sigBG}-\ref{fig:set3-sigBG} show the distribution
of event rates for background.  The efficiency of the event selection
criteria for rejecting background can be obtained from
Fig.~\ref{fig:efficiency} (top), where the simulated background and
30\% of the data sample match well at each cut level.

\begin{table*}
  \caption{Predicted signal event rates with statistical error after application of each 
    set of selection criteria. LP, ILP and DB represent lollipop, 
    inverted lollipop and double-bang, respectively. 
    For signal rates, the flux was normalized to the WB bound. 
    (The first column provides labels for 
    reference purposes in subsequent tables and figures.)}
\label{tab:rate-sig} 
\begin{tabular}{|c|c|c|c|c|c|c|c|}
\hline
 &  & \multicolumn{6}{|c|}{MC simulation}       \\
Set No. & Selection Criteria & \multicolumn{6}{|c|}{Signal $\nu$ (E$^{-2}$)}  \\
          & & LP & ILP Â&DB & $\nu_{\tau}$ & $\nu_{\mu}$ & $\nu_{e}$ \\
          & & $\times$ 10$^{-9}$ [Hz] & $\times$ 10$^{-9}$ [Hz] & $\times$ 10$^{-9}$ [Hz] & $\times$ 10$^{-8}$ [Hz] & $\times$ 10$^{-8}$ [Hz] & $\times$ 10$^{-8}$ [Hz] \\ 
\hline
EHE & NDOM $>$ 80 & 3.48 $\pm$ 0.11 & 3.54 $\pm$ 0.09 & 4.45 $\pm$ 0.16 & 50.5 $\pm$ 0.5 & 119 $\pm$ 2.2 & 39.9 $\pm$ 0.7\\ 
S1  &\lqdenI,\lqdenIII$\,>5\,$p.e./ns & 3.42 $\pm$ 0.11 & 3.05 $\pm$ 0.08 & 4.30 $\pm$ 0.16 & 24.0 $\pm$ 0.3 & 29.3 $\pm$ 0.8 & 23.9$\pm$ 0.6 \\ 
S2  & $\bar{Z}_{\rm init} < 450$~m, $\bar{V}_{Z} < -0.1$ m/ns & 2.55 $\pm$ 0.10 & 2.91 $\pm$ 0.08 & 3.95 $\pm$ 0.16 & 22.6 $\pm$ 0.3 & 24.9 $\pm$ 0.8 & 22.9 $\pm$ 0.5 \\
S3  & $Z_{\rm cog} >-330$~m, ToI $> 0.1$ & 2.32 $\pm$ 0.10 & 2.29 $\pm$ 0.08 & 3.02 $\pm$ 0.14 & 15.7 $\pm$ 0.3 & 11.8 $\pm$ 0.6 & 17.5 $\pm$ 0.5 \\
S4  & \irmax $\geq 200$, \lnpe $\geq 4.2$& 1.72 $\pm$ 0.08 & 1.72 $\pm$ 0.06 & 2.07 $\pm$ 0.11 & 5.63 $\pm$ 0.08 & 3.70 $\pm$ 0.15 & 9.08 $\pm$ 0.2 \\
\hline
\end{tabular}
\end{table*}
\begin{table*} %[H]
  \caption{\label{tab:rate-bg} Predicted background event rates with statistical error 
    after application of each set of selection criteria.  For     
    conventional neutrinos (labeled ``conv'' in the table), the 
    Bartol model~\cite{Bartol} was used. 
    For prompt neutrinos, the Martin GBW model~\cite{Martin-GBW} was used for 
    $\nu_{\tau}$ and the Sarcevic standard model~\cite{promptNu-2} was used
    for $\nu_{\mu}$  and $\nu_{e}$. }
\begin{tabular}{|c|c|c|c|c|c|c|c|} % 8 columns
\hline
Set No. & \multicolumn{6}{|c|}{MC simulation}                                          & Data \\
          & \multicolumn{5}{|c|}{Background $\nu$} & Background $\mu$ & 30 $\%$ sample \\
          & $\nu_{\mu}^{\rm conv}$ & $\nu_{e}^{\rm conv}$ & $\nu_{\tau}^{\rm prompt}$ & $\nu_{\mu}^{\rm prompt}$ & $\nu_{e}^{\rm prompt}$ &  &  \\
          & $\times$ 10$^{-8}$ [Hz] & $\times$ 10$^{-8}$ [Hz] & $\times$ 10$^{-10}$ [Hz] & $\times$ 10$^{-8}$ [Hz] 
             & $\times$ 10$^{-8}$ [Hz]& $\times$ 10$^{-6}$ [Hz] & $\times$ 10$^{-6}$ [Hz] \\
\hline
EHE &  184 $\pm$ 14.0 & 6.88 $\pm$ 0.26 & 33.4 $\pm$ 0.4  & 23.6 $\pm$ 0.50  & 9.95 $\pm$ 0.13  & 830,000 
          & 1,370,000 $\pm$ 438 \\
S1  & 8.21 $\pm$ 1.80 & 0.96 $\pm$ 0.06 & 9.74 $\pm$ 0.17 & 2.19 $\pm$ 0.12 & 3.46 $\pm$ 0.05 & 303     & 246 $\pm$ 5.9 \\
S2  & 8.11 $\pm$ 1.80 & 0.96 $\pm$ 0.06 & 9.62 $\pm$ 0.17 & 2.05 $\pm$ 0.12 & 3.42 $\pm$ 0.05 & 41.2    & 53.3 $\pm$ 2.7 \\
S3  & 4.16 $\pm$ 0.66 & 0.70 $\pm$ 0.06 & 7.12 $\pm$ 0.14 & 1.26 $\pm$ 0.09 & 2.55 $\pm$ 0.04 & 14.4    & 20.8 $\pm$ 1.7 \\
S4  & 0.24 $\pm$ 0.06 & 0.04 $\pm$ 0.003 & 0.91 $\pm$ 0.03 & 0.15 $\pm$ 0.02 & 0.43 $\pm$ 0.01 & 0.026 $\pm$ 0.01 & 0 \\
\hline
\end{tabular}
\end{table*}

\begin{figure}[h]
\includegraphics[width=86mm]{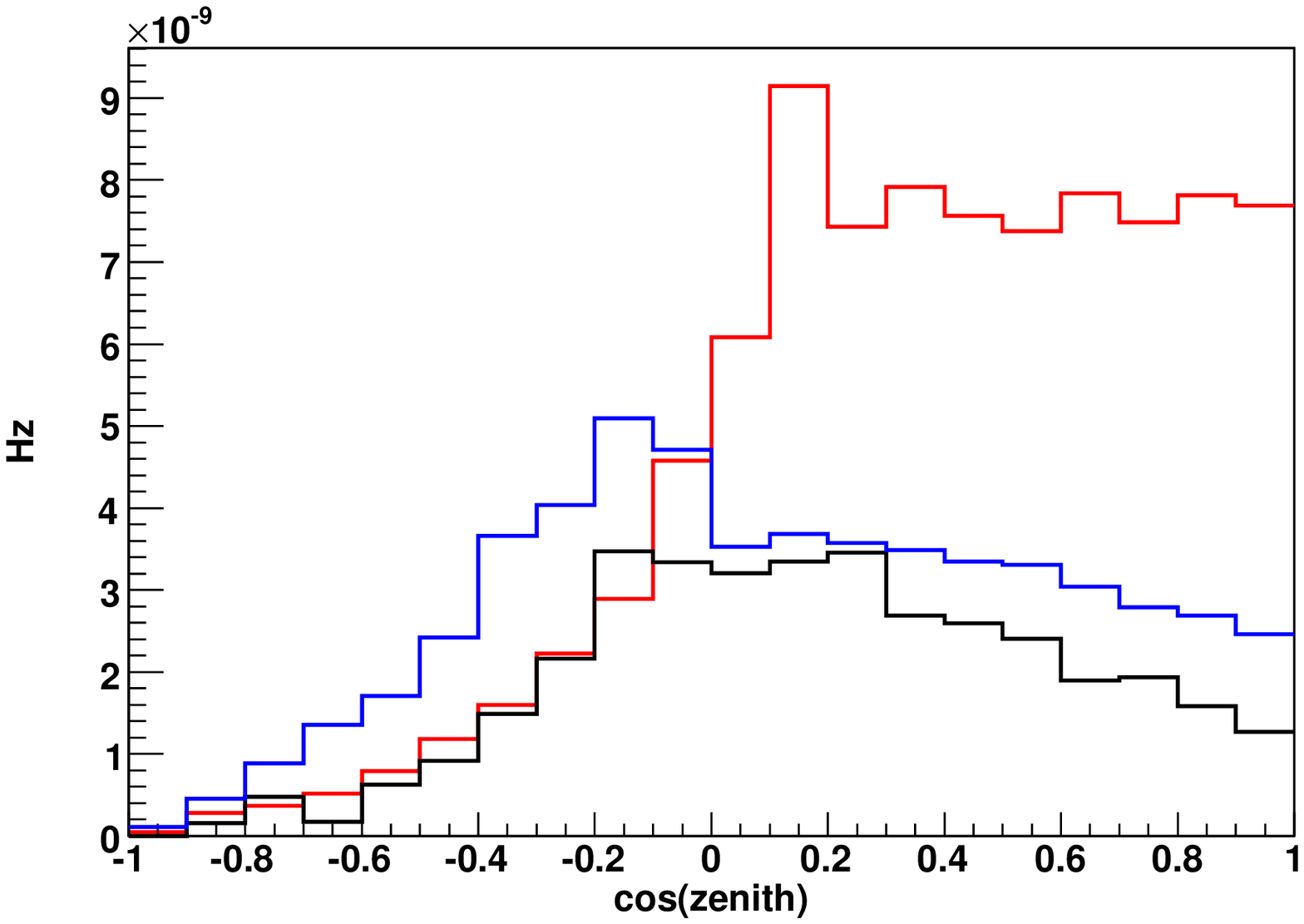}\\
\includegraphics[width=86mm]{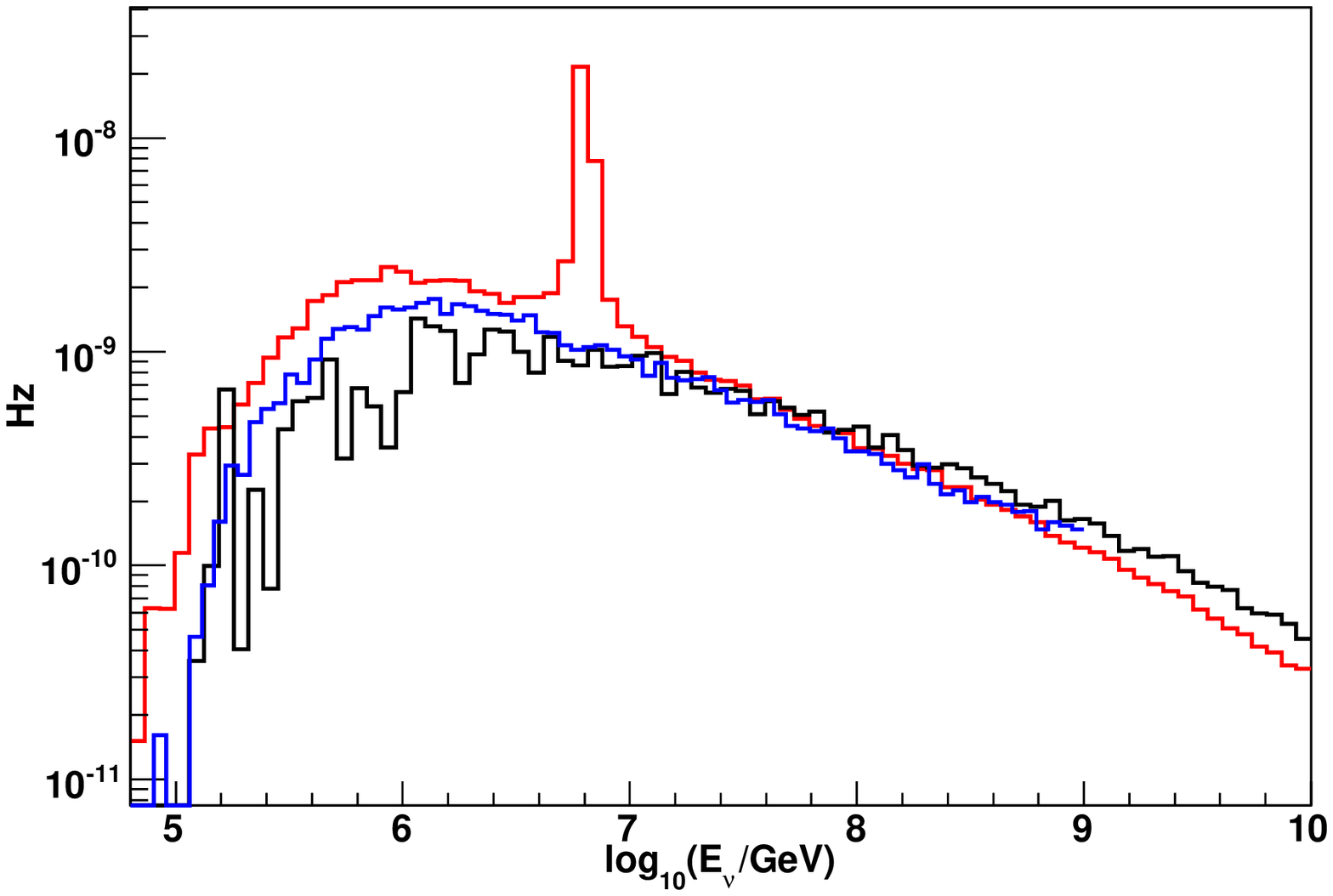}
\caption{\label{fig:mcZen} (Color online) The distribution of the true
  zenith angle (top) and the primary neutrino energy (bottom) from the
  simulation for the events passing all selection criteria. The primary 
  neutrino energies in the x-axis were weighted to E$^{-2}$ spectrum.
  Red, blue, and black lines correspond to \nueNS, \nutauNS, and \numuNS,
  respectively.}
\end{figure}

Figure~\ref{fig:mcZen} shows the distributions of the true zenith angle (top)
and primary neutrino energy (bottom) from the simulation for the events that
passed all the selection criteria. As expected, most \nutau were from
near the horizon, with the angular acceptance peaking at about
100$^{\circ}$ from vertical.

%======================================
%       Results
%--------------------------------------- 
\section{Results}
\label{sec:results}

\begin{figure}[h]
\begin{center}
\subfloat{\fbox{\includegraphics[width=50mm,height=50mm]{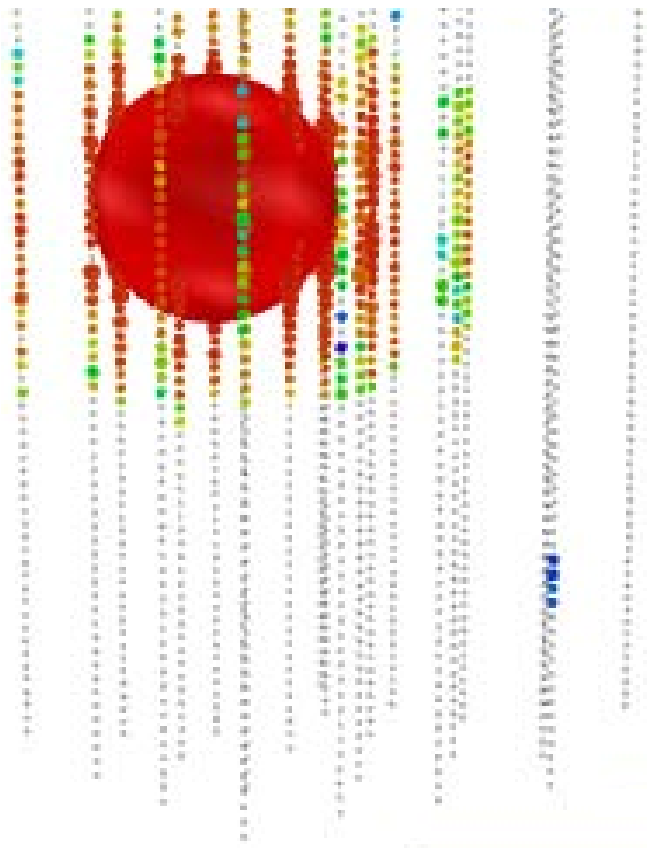}}}\\
\subfloat{\fbox{\includegraphics[width=50mm,height=50mm]{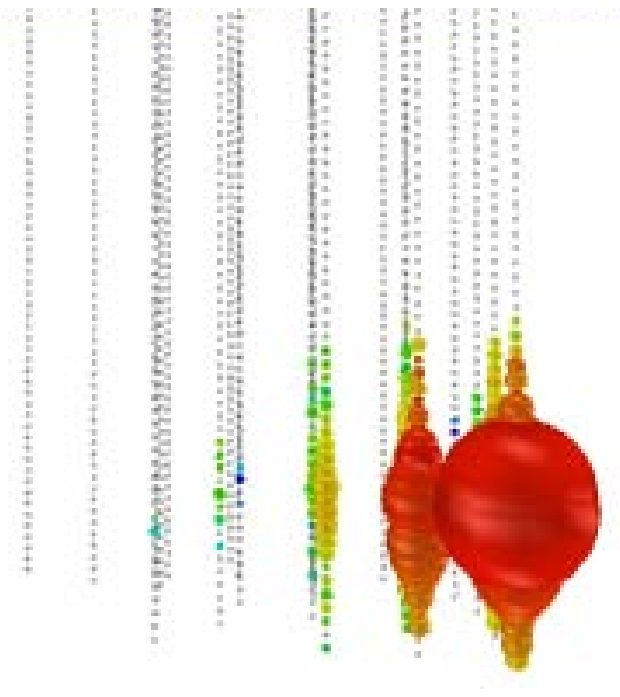}}}\\
\subfloat{\fbox{\includegraphics[width=50mm,height=50mm]{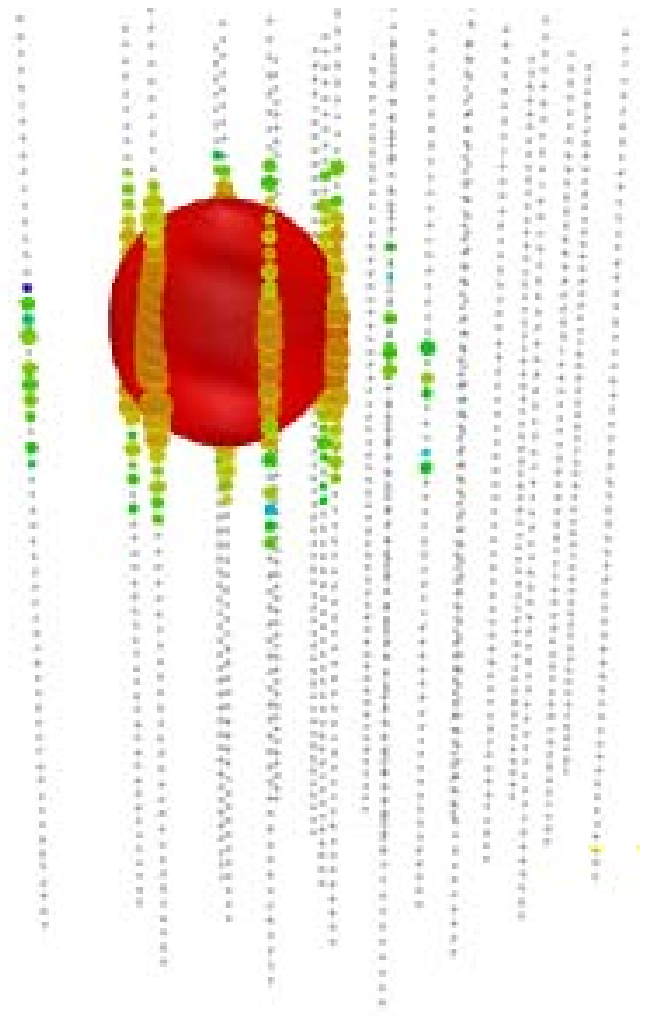}}} 
\caption{\label{fig:3events} (Color online) Diagrams of the three events surviving
  the final selection criteria applied to the 200 live-days of IC22
  data.  The radius of each circle is proportional to the number of
  photons detected by the PMT.}
\end{center}
\end{figure}

After unblinding the remaining 200 live-days of data and applying all
the selection criteria, three events remained in the data sample.  The
predicted background from all simulated sources was $0.60 \pm 0.19$ 
events.  The remaining data events are shown in
Fig.~\ref{fig:3events}.

From a detailed study of these events, we determined that one was
consistent with light produced by an AMANDA optical module observed to
emit light intermittently (Fig.~\ref{fig:3events}, top).  A second
event was qualitatively consistent with background from a nearly
horizontal muon interacting near the bottom of the detector
(Fig.~\ref{fig:3events}, middle).  The third event had the
characteristics of a neutrino-induced shower (Fig.~\ref{fig:3events},
bottom), and was also in the final sample of an independent IC22
analysis that searched for shower-like signals~\cite{IC22-cscd}.
However, we can not rule out this event as being produced by a
cosmic-ray muon accompanied by a stochastic high-energy
bremsstrahlung energy loss process.  We have conservatively included all 
three events in the final sample in the derivation of the final result.

%=======================================
%       Systematics 
%---------------------------------------
\section{Systematic and Statistical Errors}
\label{sec:systematics}

The systematic and statistical errors in this analysis were obtained
using signal and background simulations and are summarized in
Table~\ref{tab:systematics}.  In the following subsections, systematic
errors on signal and background are explained followed by our result
including both errors.

\subsection{Systematic Errors for Signal}
\label{sec:systematics-signal}

The systematic error due to our lack of precise knowledge of the DOM
sensitivity to photons was obtained by simulating the effect of
setting it to 90\% and 110\% of its nominal value resulting in
[--4.7\%, +7.9\%] error.  The systematic error in the event rates
reflecting uncertainties on the optical properties of the ice was
obtained by simulating events using different ice models.  The ice
models were created from data generated using {\it in situ} light
sources.  The baseline ice model~\cite{AHA} for this analysis used
optical properties of the ice measured at AMANDA depths and
extrapolated to IceCube depths, while an alternative ice
model~\cite{SPICE} obtained them with a direct fit to the full range
of IceCube light source data.  Comparing the predictions of the two
ice models resulted in a +29.4\% error.

The systematic uncertainty in the neutrino cross section came from two
sources.  One was from theoretical uncertainty in the parton
distribution function evaluation and structure function and the other
was from errors in the experimental measurement of the parton
distribution function by HERA~\cite{HERA}.  From these two sources we
estimated the systematic error in the neutrino cross section as $\pm
6.4$\%. Very high energy events could saturate PMTs by exceeding the
PMT's dynamic range. This could result in an incorrect estimation of
the original neutrino energy.  Since the observable quantity most
closely related to the energy is \npe, the systematic error
associated with the PMT saturation was obtained by observing the
impact of changing the \npe cut from 90\% to 110\% of its original
value.  This error was found to be [--5.7\%, +5.0\%].

\subsection{Systematic Errors for Background}
\label{sec:systematics-bg}

The systematic errors due to uncertainties in DOM sensitivity, ice
properties, and DOM saturation behavior were obtained in the same
manner as for the signals, as described in
Section~\ref{sec:systematics-signal}.  They were estimated as
[--4.7\%, +7.9\%], [--62\%, +85\%], and [--28.9\%, +5.3\%],
respectively.

In addition, there are systematic errors which applied only to the
background.  The muon event rate is known to change as a function of
the atmospheric temperature above the South Pole plateau.  Since our
muon simulation assumed a rate pegged to that seen in October, the
seasonal variation was taken into account as a systematic error and
was estimated as [--24\%, +18\%] when compared with IC22 data at
EHE filter level.  The systematic error due to cosmic ray composition
was also obtained by switching constants and slopes between proton and
iron in the two component model data.  At S3, just before the final
cut to have enough statistics, we obtained --24\% by this method.

There are alternative models for the prompt neutrino flux.  For this
analysis, the base models used for the prompt neutrino flux are
Sarcevic standard flux model for \numu and \nue~\cite{promptNu-2}, and
Martin GBW model for \nutau~\cite{Martin-GBW}.  As an alternative, we
have also considered the Sarcevic minimum and maximum flux
models~\cite{promptNu-2}, from which we estimate a [--59\%, +30\%]
systematic error on the prompt neutrino flux.

\begin{table}[H]
  \caption{\label{tab:systematics} Summary of the systematic and 
    statistical errors for signal and background events from the simulated data.}
  \begin{tabular}{|c|c|c|}
    \hline
    Source & Signal & Background \\
    \hline
    DOM sensitivity & -- 4.7$\%$, + 7.9$\%$ &  -- 4.7$\%$, + 7.9$\%$  \\
    Ice properties  & -- 0$\%$, + 29$\%$ & -- 62$\%$, + 85$\%$ \\
    $\nu$ cross section  & -- 6.4$\%$, + 6.4$\%$ & N/A \\
    PMT saturation  & -- 5.7$\%$, + 5.0$\%$ & -- 29$\%$, + 5.3$\%$ \\
    Cosmic ray flux & N/A & -- 0$\%$, + 16$\%$ \\ 
    Cosmic ray composition & N/A  & -- 24$\%$, + 0$\%$ \\
    Seasonal variation & N/A &  -- 24$\%$, + 18$\%$ \\
    Prompt $\nu$ flux model & N/A  & -- 59$\%$, + 30$\%$  \\
    \hline 
    Total Syst. error & -- 7.9$\%$, + 31$\%$ & -- 97$\%$, + 94$\%$ \\ 
    \hline 
    Total Stat. error & $\pm 2.3 \%$ & $\pm 32 \%$ \\
    \hline
  \end{tabular} 
\end{table} 

\subsection{Result including Statistical and Systematic Errors}

Since it was computationally feasible to generate a large amount of
simulated signal, the statistical error on the simulated signal is
small ($\pm 2.3$\%).  By contrast, the considerably larger statistical
error on the simulated background ($\pm 32$\%) reflects the aggregate
effect of the high rejection efficiency of our selection criteria and
the limitations imposed by finite computational resources.  In
summary, the expected signal and background events for 200 live-days
with IC22 are $3.18 \pm 0.07$ (stat.) $^{+2.99}_{-3.08} $ (syst.) and
$0.60 \pm 0.19$ (stat.) $^{+0.56}_{-0.58}$ (syst.), respectively.
When we unblinded 200 live-days of data we observed 3 events which
were deemed compatible with background.  With a predicted background
of $0.60 \pm 0.19$ (stat.)  $^{+0.56}_{-0.58}$ (syst.) events, the
probabilities of observing one, two or three events due solely to
fluctuations in the background are 30\%, 13\% and 5\%,
respectively.

We combined the systematic errors in quadrature with the statistical
errors and applied a profile log-likelihood method~\cite{Trolke} to
obtain the confidence interval~\cite{FeldmanCousins}.  The 90\%
confidence level (CL) upper limit on signal for 200 live-days was
obtained as $\mu_{90}^{\text{s}} = 7.7$ events.  The 90\% CL upper
limit on astrophysical all-flavor neutrino flux, $\Phi_{90}(\nu_{\rm
  x})$, was obtained using the following relation:
$\frac{\Phi_{90}}{\Phi_{\text{WB}}} =
\frac{\mu_{90}^{\text{s}}}{\text{N}_{\text{WB}}}$ where
$\Phi_{\text{WB}}$ and $\text{N}_{\text{WB}}$ are the WB bound for
all-flavor astrophysical neutrinos and the corresponding number of
all-flavor astrophysical neutrinos for 200 live-days, respectively.
The obtained 90\% CL upper limit is \theLimit for the 3 observed
events from the 200 live-days of IC22 data.

\begin{figure}[b]
\includegraphics[width=86mm, height=70mm]{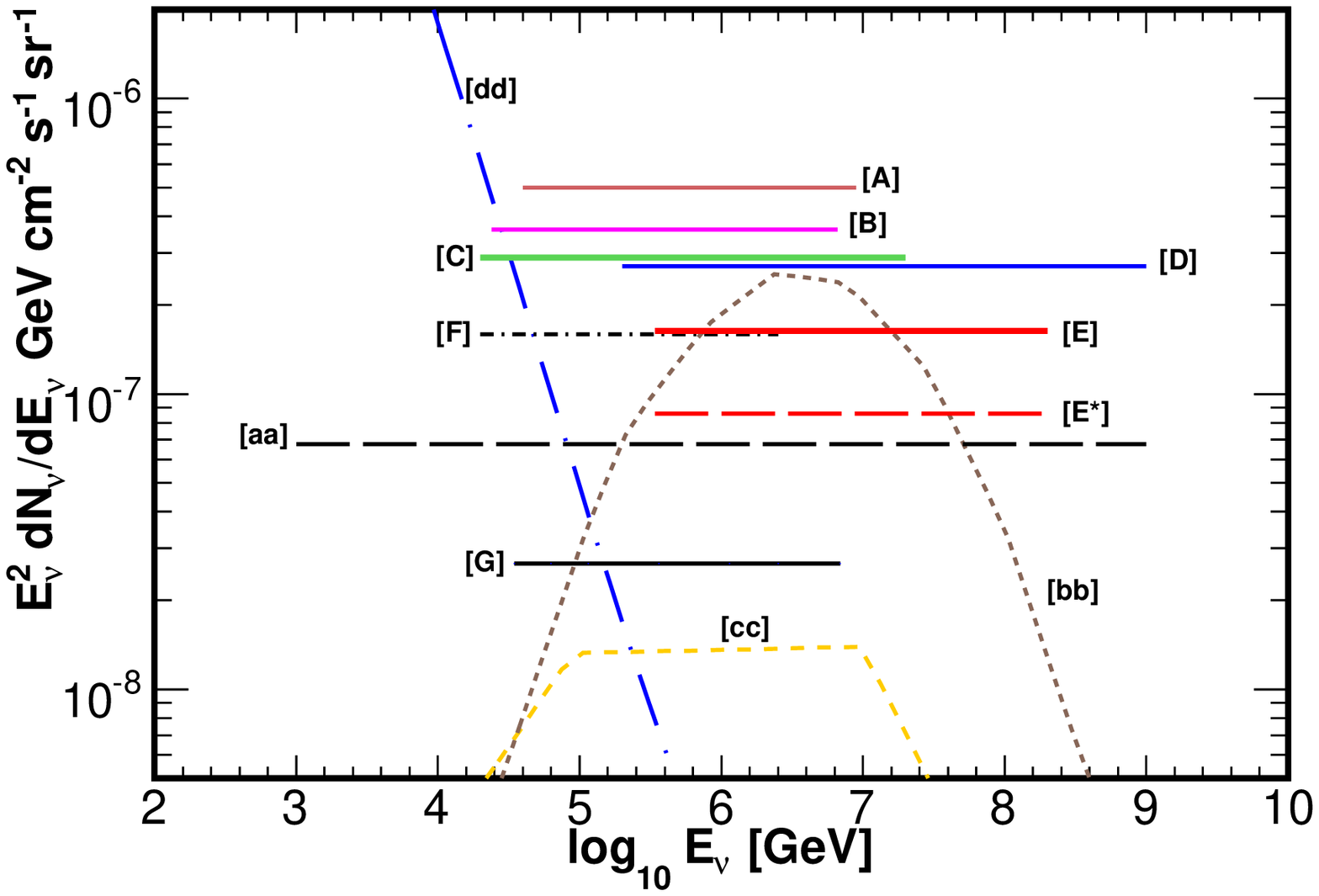}
\caption{\label{fig:limit} (Color online) The limits on production of UHE neutrinos.
  [A]: AMANDA-II cascade all-flavor limit (1001 live-days)~\cite{AMANDA-II-cscd}, 
  [B]: IC22 cascade all-flavor limit (257 live-days)~\cite{IC22-cscd}, 
  [C]: Baikal all-flavor limit (1038 live-days)~\cite{Baikal}, 
  [D]: AMANDA-II UHE all-flavor limit (457 live-days)~\cite{AMANDA-II-UHE}, 
  [E]: IC22 UHE all-flavor limit from diffuse
  sources using the analysis described in this paper (this work, 200 live-days), 
  [E*]: IC22 UHE all-flavor sensitivity (this work, 200 live-days),
  [F]: ANTARES '07-'09 \numu x 3 334 d~\cite{ANTARES-neutrino2010},
%  [F]: IC40 UHE all-flavor limit (345.7 live-days, preliminary)~\cite{IC40-UHE-all}, 
  [G]: IC40 \numu x 3  (375.5 live-days)~\cite{IC40-Sean},
  [aa]: Waxman-Bahcall (\numu and \numubar) model 1998 $\times \frac{3}{2}$~\cite{WB}, 
  [bb]: Stecker AGN (Seyfert) 2005~\cite{Stecker-2005},
  [cc]: Waxman-Bahcall Prompt GRB model~\cite{WB}, and
  [dd]: Atmospheric neutrino flux (Bartol + Sarcevic standard model)
}

\end{figure}

This limit applies to the primary neutrino energy range of $340\,{\rm
  TeV} < E_\nu < 200\,{\rm PeV}$, covering the middle 90\% of the
accepted simulated signal.  Fig.~\ref{fig:limit} shows this limit
together with several theoretical model predictions.  The upper limit
on the tau neutrino flux is one third that of the all-flavor
astrophysical neutrino flux if one assumes a flavor ratio of
\nueNS:\numuNS:\nutau = 1:1:1 at Earth.

%====================================== 
%        Summary and Outlook
%---------------------------------------
\section{Conclusions and Outlook}
\label{sec:conclusions}

A set of selection criteria designed for UHE \nutau detection were
applied to IceCube data.  These criteria also had appreciable
efficiency for UHE \nue and \numu detection.  We applied these
criteria to 200 live-days of data from IceCube's 22-string
configuration and observed 3 events in the final sample. We
therefore set a 90\% CL upper limit on the astrophysical UHE
all-flavor neutrino flux of \theLimit.  The analysis improves on the
previous limit set by AMANDA~\cite{AMANDA-II-cscd, AMANDA-II-UHE,
  IC22-cscd} with comparable integrated exposure.  Future IceCube
searches specialized for \nutau will be more sensitive due to the
increased instrumented volume relative to IC22.  The large volume will
also warrant the application of sophisticated \nutau reconstructions,
further improving the sensitivity of these searches.

\vspace{0.2cm}

\begin{acknowledgments}     
We acknowledge the support from the following agencies:
U.S. National Science Foundation-Office of Polar Programs,
U.S. National Science Foundation-Physics Division,
University of Wisconsin Alumni Research Foundation,
the Grid Laboratory Of Wisconsin (GLOW) grid infrastructure at the University of Wisconsin - Madison, the Open Science Grid (OSG) grid infrastructure;
U.S. Department of Energy, and National Energy Research Scientific Computing Center,
the Louisiana Optical Network Initiative (LONI) grid computing resources;
National Science and Engineering Research Council of Canada;
Swedish Research Council,
Swedish Polar Research Secretariat,
Swedish National Infrastructure for Computing (SNIC),
and Knut and Alice Wallenberg Foundation, Sweden;
German Ministry for Education and Research (BMBF),
Deutsche Forschungsgemeinschaft (DFG),
Research Department of Plasmas with Complex Interactions (Bochum), Germany;
Fund for Scientific Research (FNRS-FWO),
FWO Odysseus programme,
Flanders Institute to encourage scientific and technological research in industry (IWT),
Belgian Federal Science Policy Office (Belspo);
University of Oxford, United Kingdom;
Marsden Fund, New Zealand;
Australian Research Council;
Japan Society for Promotion of Science (JSPS);
the Swiss National Science Foundation (SNSF), Switzerland.
\end{acknowledgments}

\clearpage

\bibliography{basename of .bib file}

\end{document}